\documentclass[12pt,final]{iopart}
\usepackage{graphicx,amssymb,bbm}

\def\ii{ {\rm i} }
\def\dd{ {\rm d} }

\newcommand{\Bra}[1]{{( #1 |}}
\newcommand{\Ket}[1]{{| #1 )}}
\def\Braket#1#2{( #1 | #2 )}

\def\BBraket#1#2{(\!( #1 | #2 )\!)}
\newcommand{\ave}[1]{{\langle #1 \rangle}}

\def\aT{{\mathbf{T}}}
\def\aP{{\mathbf{P}}}
\def\aF{{\mathbf{F}}}
\def\bT{\hat{\mathbf{T}}}
\def\cT{\tilde{\mathbf{T}}}
\def\cK{\tilde{\mathbf{K}}}
\def\aS{{\mathbf{S}}}
\def\aW{{\mathbf{W}}}
\def\bP{\hat{\mathbf{P}}}
\def\bR{\hat{\mathbf{R}}}
\def\C{{\mathbb{C}}}
\def\Z{{\mathbb{Z}}}

\def\one{\mathbbm{1}}

\def\>{\rangle}
\def\<{\langle}
\def\ket#1{|#1\>}
\def\bra#1{\<#1|}
\def\braket#1#2{\< #1 | #2 \>}

\def\ave#1{\< #1\>}
\def\ve#1{{\underline{#1}}}

\def\tr{{{\rm tr}}}

\def\re{{\,{\rm Re}\,}}

\newcommand{\x}{{\mathrm{x}}}
\newcommand{\y}{{\mathrm{y}}}
\newcommand{\z}{{\mathrm{z}}}
\newcommand{\s}[2]{\sigma^{#1}_{#2}}

\begin{document}

\title{Chaos and Complexity of quantum motion} 

\author{Toma\v z Prosen}
\address
{Physics Department, Faculty of mathematics and physics, University of Ljubljana, 
Jadranska 19, SI-1000 Ljubljana, Slovenia}
    
 \date{\today}

\begin{abstract}
The problem of characterizing complexity of quantum dynamics - in particular of locally interacting chains of quantum
particles - will be reviewed and discussed from several different perspectives: (i) stability of motion against external perturbations and decoherence, (ii) efficiency of quantum simulation in terms of classical computation and entanglement production in 
operator spaces, (iii) quantum transport, relaxation to equilibrium and quantum mixing, and (iv) computation of quantum 
dynamical entropies. Discussions of all these criteria will be confronted with the established criteria of integrability or 
quantum chaos, and sometimes quite surprising conclusions are found. Some conjectures and interesting open problems 
in ergodic theory of the quantum many problem are suggested.
\end{abstract}


 \bigskip
\noindent
{\em Invited review article for special issue of Journal of Physics A on Quantum Information}

\section{Introduction}
This article will be about discussing the possibilities to characterize {\em randomness} 
and {\em dynamical complexity} in quantum mechanics and relating this issue to the questions of 
non-equilibrium statistical mechanics.  We shall try to illustrate, mainly by presenting various numerical examples, a possible `cyclist approach' \cite{cvitanovic} towards the {\em quantum many-body problem} which is inspired by experiences gained in studies of quantum and classical chaos 
of one or few particles (see e.g. \cite{Haake:00,cvitanovic}).

Solving the many body problem in quantum mechanics presents a major challenge from its very beginnings. And along this way, many ingenious important analytical and efficient numerical methods of solution have been developed, for example Bethe ansatz \cite{bethe}, later generalized and interpreted as the quantum inverse scattering problem \cite{korepin},  real space 
renormalization group methods \cite{wilson},  quantum Monte Carlo \cite{QMC}, density matrix renormalization group (DMRG) \cite{white}, and more recently various quantum information-theoretic based time-dependent DMRG (tDMRG) \cite{MPS,SchWhite,tdmrg}. The ultimate aim of any of these methods is to efficiently find analytical or numerical approximation to the solution for some of the
physical observables in the quantum many body problem, however many methods work only under
some specific conditions which are not always well understood.
For example, Bethe ansatz and quantum inverse scattering work only for a small
subset of problems which are {\em completely integrable}, and which may often have very 
non-generic non-equilibrium properties, such as e.g. ballistic transport at high temperatures \cite{Zotos}.
Quantum Monte Carlo techniques represent a very successful
set of numerical methods which can yield thermodynamic equilibrium averages for generic (non-integrable) systems, however they are practically useless for computing non-equilibrium, or time dependent quantities, such as transport coefficients.
Traditional DMRG method \cite{white} is provably very successful for computing accurate ground state 
expectation values of almost any physical observables of one-dimensional interacting systems. And tDMRG 
\cite{MPS,tdmrg} promised to extend the
success of DMRG to time dependent physics. Indeed, the first numerical experiments looked very promising, but after a closer look one may realize that they have all been applied to a rather special subset of interacting systems and to a rather special subset of initial states. 
For generic (non-integrable) interacting systems or for sufficiently complex initial states, tDMRG should fail to provide an efficient computation as shall be discussed below.
In our view this is to be expected, and represents an
intrinsic characteristic of quantum complexity of such system and 
 should correspond to many body extensions of the phenomena of quantum chaos \cite{Haake:00} where spectra and eigenvector coefficients can be described by statistical ensembles of random matrices.
In the past two, almost three decades there has been a lot of activity in the so called field of quantum chaos, or quantum chaology \cite{berry}, where people were trying to understand the essential and significant features of quantum systems which behave chaotically in the classical
limit. Classical chaos can be defined in terms of positive (algorithmic, Kolmogorov-Chaitin) complexity of systems' orbits.
Still, the question whether such definition of complexity can be in an intuitively sensible way translated to quantum mechanics failed to be answered in spite of many efforts. 
It seems that quantum systems of finite (one or few body) chaotic systems are generically more robust against imperfections \cite{fidelityreview} than their classical counterparts, which is consistent with a simple
illustration based on wave-stability of unitary quantum time evolution \cite{casati86}. 

Nevertheless, it has been suggested that exponential sensitivity to initial conditions, the essential characteristics of classically chaotic systems, has many fingerprints
in quantum mechanics. For example, in one of the pioneering works on quantum Loschmidt echoes (or fidelity decay), Jalabert and Pastawski \cite{jalabert}
have shown that in the semi-classical regime, decay of system's sensitivity to external perturbation, as defined by fidelity, is exponential
with the rate which can be related to classical Lyapunov exponents. However, this is only true in sufficiently semi-classical regime, where effective
Planck constant is smaller than effective strength of perturbation, and where fidelity can be essentially explained classically \cite{veble}. On the other hand,
in purely quantum regimes, quantum fidelity decays in completely different manner than the classical fidelity, and quite surprisingly displays slower decay for systems with stronger decay of temporal correlations \cite{fidelityreview}.

Throughout this paper we shall discuss various ergodic properties of a simple toy model of non-integrable interacting  quantum many body system, namely kicked Ising (KI) chain \cite{ProsenPRE02},
and its time-independent version,  which (both) undergo a transition from integrable to quantum chaotic regimes when the direction of the (kicking) magnetic field is changed. By simulating the dynamics of the model in terms of tDMRG we find that 
it can be performed efficiently only in the integrable regimes. 
Being interested in statistical mechanics of the model we shall describe numerical experiments addressing the questions on the relationship between the onset of quantum chaos in KI model and its quantum mixing property, namely the nature of relaxation to equilibrium, and the properties of non-equilibrium steady state.
On one hand we argue that the regime of quantum chaos essentially corresponds to the regime of quantum ergodicity and
quantum mixing where diffusive transport laws are valid.
On the other hand, we conjecture that the transition to non-ergodic regime may occur before the system parameters
reach the integrable point (even in thermodynamic limit), and that non-ergodic to ergodic transition can 
be characterized with order parameters which change discontinuously at a critical value of system parameter.
We argue that the process of relaxation to equilibrium in quantum chaotic (mixing) case can be described
in terms of quantum analogues of Ruelle resonances \cite{Prosen04}.
Furthermore, we conjecture that the eigenvectors of this process possess a certain scale invariance which can be
described by simple power laws. 
We also discuss a possibility of numerical calculation of quantum dynamical entropies 
\cite{CNT,AF,Benatti} 
in a non-trivial setting of KI model, and find, quite remarkably, that positivity of such dynamical entropies
{\em does not correspond} to any other measures of quantum chaos, namely quantum dynamical entropies appear
to be positive even in the integrable regimes.
  
About  two thirds of the material presented in this paper constitutes a review of a selection of recent results related to quantum chaos in many-body systems, with a flavor of quantum information.  However, about one third of the
material, constituting a major part of section \ref{sect:statmech}, is new and original and has not yet been published before.
The paper is organized as follows. In short section \ref{sect:definitions} we review basic definitions
of complete integrability and quantum chaos, in particular in the context of many-body systems.
In section \ref{sect:model} we introduce KI model
 which will serve us as a very convenient and efficient toy model to numerically demonstrate all the phenomena discussed in this paper.    In section \ref{sect:losch} we discuss the robustness of quantum systems
to external perturbations and decoherence in open quantum systems, mainly as characterized by quantum
Loschmidt echoes and entanglement between the system and the environment.
In section \ref{sect:MPS} we discuss the time efficiency of best (known) classical simulation, namely
using tDMRG, of locally interacting 1D quantum systems and its possible dependence on integrability of the model.
In section \ref{sect:transport} we relate standard criteria of quantum chaos to normal (diffusive) versus anomalous (ballistic) transport and discuss a simulation of a quantum heat current in non-equilibrium steady state.
In section \ref{sect:statmech}  we discuss a problem of quantum relaxation to equilibrium, i.e. the quantum mixing
property, and quantitative characterization of quantum dynamical complexity.
This section contains a large portion of original intriguing numerical results which support
few interesting and perhaps surprising conjectures.
In section \ref{sect:conc} we conclude and discuss some open problems.

\section{Integrability versus chaos}

\label{sect:definitions}

The central issue of this paper is to verify and demonstrate to what extend the 
{\em complete integrability} affects non-equilibrium properties 
of quantum many-body systems and their dynamical complexities, 
and conversely, whether {\em (strong) non-integrability} 
generically coincides with the established criteria of {\em quantum chaos}.

Let us start by giving some established definitions 
(see e.g. \cite{cvitanovic,Haake:00,solitons,korepin})
of the basic terms needed to understand the issue.

A classical Hamiltonian many-body system with $L$ degrees of freedom is {\em completely integrable}, if there exist $L$ functionally independent global smooth phase space functions (integrals of motion) which are mutually 
in involution, i.e. all Poisson brackets among them vanish.
In such a case global canonical transformation to canonical {\em action-angle} variables
can be constructed, and dynamics can be explicitly solved - at least in principle.
For locally interacting {\em infinite} systems ($L\to\infty$) analytic methods for an explicit construction of integrals of motion and canonical action-angle variables are known 
which usually go under a common name of {\em inverse scattering method}.
Explicit solution by a mapping to an iso-spectral Sch\" rodinger problem in terms of inverse scattering
technique is usually understood as a definition of {\em complete integrability} in such context.

Definition of {\em quantum complete integrability} is less unique. Nevertheless,
algebraic, non-commutative versions of the inverse scattering technique exist and can be applied to some quantum interacting lattices in one-dimension, generalizing
the famous Bethe ansatz solution of the Heisenberg spin $1/2$ chain,
and this is used as the most general known definition of 
{\em quantum complete integrability}. Other versions of integrability for finite $L$ quantum
systems have been proposed but they do not correspond to the integrability of
the underlying classical limit, if the latter exists, so these ideas will not be considered in this paper.

It has not been generally accepted yet, though demonstrated in many occasions, that completely
integrable quantum systems constitute only a small subset of physical models and posses many exceptional (non-generic) non-equilibrium dynamical properties, like for example anomalous transport at finite temperatures (see e.g.\cite{Zotos}.)

On the other extreme of ergodic hierarchy we have chaotic systems.
In classical Hamiltonian dynamics of few particles, chaos is best defined in terms of positive
algorithmic (Kolmogorov) complexity of systems' trajectories, or equivalently, by exponential 
sensitivity to initial conditions. 
However, bounded quantum systems of finite number of particles cannot be dynamically complex
as their excitation spectrum is discrete, and hence the evolution is necessarily quasi-periodic 
(or almost periodic). Still, quite surprisingly, even for such systems certain dynamical properties 
are {\em random} and {\em universal}, if the underlying classical limit is sufficiently strongly 
chaotic \cite{berry,Haake:00}.
But genuine dynamical complexity may emerge in thermodynamic limit.
However, there is still no completely satisfactory definition of dynamical chaos for infinite quantum systems.
The commonly used working definition is the reference with the {\em random matrix theory} 
\cite{Haake:00}, namely the many-body quantum system is said to be quantum chaotic if its excitation
spectrum or some other dynamical properties can be (on certain energy, or time scale)
well described by ensembles of random Hermitian matrices with appropriate symmetry
properties. 

\section{Toy model}

\label{sect:model}

Throughout this paper we shall use, either for illustration of theoretical results, or for numerical experimental studies, 
the following 1D locally interacting quantum lattice system, namely a chain of $L$ qubits, or spin $1/2$ particles, 
coupled with nearest neighbour Ising interaction
and periodically kicked with spatially homogeneous, but {\em arbitrarily oriented} magnetic field.
In a suitable dimensionless units our model can be written in terms of a three-parameter
periodic time-dependent Hamiltonian
\begin{equation}
H(J,h_\x,h_\z;t) = \sum_{j=0}^{L-1}\left\{ J \s{\x}{j}\s{\x}{j+1} + \delta_{\rm p}(t)(h_\x \s{\x}{j} + 
h_\z \s{\z}{j})\right\}.
\label{eq:hamiltonian}
\end{equation}
$\delta_{\rm p} = \sum_{m\in\Z} \delta(t-m)$ is a unit-periodic Dirac delta and
$\s{\x,\y,\z}{j}$ are the usual Pauli spin variables on a finite lattice $j\in\Z_L=\{0,1,\ldots L-1\}$,
satisfying the commutation (Lie) algebra 
$[\s{\alpha}{j},\s{\beta}{k}]=\sum_\gamma 2\ii\varepsilon_{\alpha \beta\gamma} \delta_{jk} \s{\gamma}{j}$.
Sometimes it will turn fruitful if we extend the set of Pauli matrices by identity matrix and assign them a numerical
superscript $\s{\alpha}{j},\alpha\in \Z_4$, namely
$\s{0}{j}\equiv \one, \s{1}{j}\equiv \s{\x}{j}, \s{2}{j}\equiv\s{y}{j}, \s{3}{j}\equiv\s{z}{j}$.
The finite chain will often be treated with {\em periodic boundary conditions}, $\s{\alpha}{L}\equiv \s{\alpha}{0}$,
and sometimes the thermodynamic limit (TL) $L\to\infty$ will be considered, in particular when we shall
study the statistical mechanics of (\ref{eq:hamiltonian}) in section \ref{sect:statmech}.

Although kicked Hamiltonian one-particle models have been very popular in the field of nonlinear
dynamics and quantum chaos for decades, for example the Chirikov's kicked rotator model \cite{chirikov}, the
use of kicked systems in quantum many body physics has been so far very limited.
If one is not only interested in zero temperature (ground state) or low temperature physics, then as we shall try to demonstrate in this review, kicked many-body models like (\ref{eq:hamiltonian}) provide
simpler and clearer phenomenological picture of global dynamics than time-independent models. The main reason is that since energy is not
a conserved quantity, the entire Hilbert space of many-body configurations is accessible
for non-integrable dynamics, and the notions of quantum ergodicity and mixing 
(see e.g. Ref.\cite{ProsenPRE99} for definitions and further references) 
can be defined more clearly than in the time-independent, autonomous setting.
Perhaps the first kicked interacting quantum lattice has been introduced in Ref.\cite{ProsenPRL98}.
Even if in traditional solid state physics such kicked dynamics would represent un-physically strong excitations, one
has to realize that kicked quantum chains could be attractive options as benchmark models of 
quantum state manipulation and quantum computation in optical lattices.

The {\em Kicked Ising} (KI) model (\ref{eq:hamiltonian}) has been defined for the first time
in Ref.\cite{ProsenPRE02},  generalizing the integrable KI model with transverse field introduced
in Ref.\cite{Prosen00}. Clearly, as shown there \cite{Prosen00} for the case of {\em transverse}
field, $h_\x=0$, the time-dependent model
(\ref{eq:hamiltonian}) can be considered {\em completely integrable} since it can be solved explicitly, 
for example by Wigner-Jordan-Bogoliubov transformation to non-interacting spinless fermions, and a 
large class of its time 
correlation functions can be calculated explicitly. In addition, an infinite sequence of local conserved 
quantities (integrals of motion) can be constructed in such a case.
There is another, trivial {\em completely integrable} regime of KI model, namely the case of {\em longitudinal} field,
$h_\z=0.$
Yet another, more non-trivial {\em completely integrable} regime of KI model is found when the magnitude
of dimensionless field is a multiple of $\pi/2$, namely $h=\sqrt{h_\x^2 + h_\z^2} = n \pi/2, n\in\Z$,
since then the magnetic kick can be considered as a multiple of $\pi$ rotation, and generated by a slightly
obscure set of non-interacting spinless fermions.

However, in a general case of titled magnetic field when both components
$h_\x$ and $h_\z$ are non-vanishing, and $2 h/\pi$ is non-integer, the model is non-integrable, and
is conjectured to be not amenable to exact analytical methods. As discussed in the following sections,
non-integrable KI model can display a variety of regimes according to the criteria of quantum chaos, 
quantum ergodicity and quantum mixing. In fact, recently the spectral statistics of KI model in 
strongly non-integrable
regime has been studied in detail and random matrix behaviour has been clearly confirmed at short
energy ranges \cite{pinedaprosen07}.

Due to kicked nature of interaction, the evolution propagator of KI model for one unit period of
time (the so-called Floquet operator), starting just before the kick, can be constructed explicitly
in terms of a time-ordered product
\begin{eqnarray}
U(J,h_\x,h_\z) &=& {{\cal T}\exp}\left(-\ii\int_{0}^{1}\!\!\dd t  H(J,h_\x,h_z;t-0)\right) \\
                          &=& \exp(-\ii J\sum_j \s{\x}{j}\s{\x}{j+1})\exp(-\ii \sum_j (h_\x \s{\x}{j} + h_\z \s{\z}{j})) \label{eq:floquet} \\
                          &=& \prod_j U''_{j,j+1}(J) \prod_j U'_{j}(h_\x,h_\z) \label{eq:protocol}.                         
\end{eqnarray}
The last line suggests a simple efficient quantum protocol to simulate KI model in terms of simple
1-qubit $U'(h_\x,h_\z) = \exp(-\ii (h_\x \s{\x}{} + h_\z \s{\z}{}))$ and
2-qubit $U''(J) = \exp(-\ii J \s{\x}{} \otimes \s{\x}{})$ quantum gates.
If we write a compact Kicked Ising 2-qubit gate  
\begin{equation}
W(J,h_\x,h_\z) = U''(J) \cdot (U'(h_\x,h_\z)\otimes\one)
\label{eq:Wgate}
\end{equation}
and introduce a left-to-right ordered operator product, namely $\prod^+_j A_j \equiv \cdots A_1 A_2 A_3 \cdots $,
then we can write KI protocol as a simple string of $W-$gates
\begin{equation}
U(J,h_\x,h_\z) = \prod^+_j W_{j,j+1}(J,h_\x,h_\z).
\label{eq:protocol2}
\end{equation}

The KI model has a better known autonomous limit,
namely time-independent Ising chain in a tilted magnetic field
\begin{equation}
H'(J,h_\x,h_\z) = \lim_{\tau\to 0} \frac{1}{\tau}\overline{H(\tau J,\tau h_\x,\tau h_\z;t)} = 
\sum_j (J \s{\x}{j}\s{\x}{j+1} +  h_\x \s{\x}{j} + h_\z \s{\z}{j}),
\label{eq:hamaut}
\end{equation}
which is again non-integrable unless the field is transverse $h_\x = 0$, or longitudinal $h_\z = 0$.

\section{Decoherence and fidelity}

\label{sect:losch}

\subsection{Loschmidt echoes}

One of the central questions about the dynamics of complex quantum systems is their robustness against
small imperfections in the Hamiltonian. While it is clear that due to unitarity the quantum evolution is stable
against variation of initial states \cite{casati86}, it is not so clear how stable it is against variation of the
Hamiltonian, either being static or time-dependent -  perhaps even noisy.

Let us write the Hamiltonian as $H_\delta = H_0 + \delta V$, where $H_0$ is the unperturbed Hamiltonian,
$V$ is a Hermitian perturbation operator and $\delta$ is a small perturbation parameter.
Peres \cite{peres84} proposed the following measure of stability of quantum evolution: Let us start from
some fixed initial state $\ket{\psi}$, and write the time evolutions of this state generated with the unperturbed and
perturbed Hamiltonian, as $\ket{\psi_0(t)}=U_0(t)\ket{\psi}$ and $\ket{\psi_\delta (t)}=U_{\delta}(t)\ket{\psi}$,
respectively. Then the stability is characterized by fidelity, {\em i.e.} the squared overlap between these two states
\begin{equation}
F(t) = |\braket{\psi_0(t)}{\psi_\delta(t)}|^2 = |\bra{\psi}U^\dagger_0(t) U_\delta(t)\ket{\psi}|^2.
\label{eq:fidelity}
\end{equation}

There are two alternative interpretations of quantum fidelity: (i) One can interpret (\ref{eq:fidelity}) as a
{\em quantum Loschmidt echo}, namely the probability that the initial state $\ket{\psi}$, which is propagated forward
with perturbed evolution $U_\delta(t)$, and after that propagated backwards in time with unperturbed evolution
$U_0^\dagger(t) = U_0(-t)$, is again measured in the same (initial) state.
Alternatively (ii) is just the square modulus of expectation value of the unitary echo operator $U_0(-t)U_\delta(t)$ which
is the quantum propagator in the {\em interaction picture}.

There have been three main theoretical approaches to understanding of fidelity decay in quantum 
dynamical systems:
\subsubsection{Semi-classical approach.} 
In a seminal work Jalabert and Pastawski derived quantum Loschmidt echo for systems which posses well
defined classical limit. They have shown that under certain conditions, namely that the perturbation strength is
large enough - typically larger than appropriately scaled Planck constant, and that initial states have certain
classical interpretations - like coherent states, position states, etc, the quantum Loschmidt echo decays exponentially
\begin{equation}
F(t) \sim \exp(-\lambda t)
\end{equation}
with the rate $\lambda$ which is {\em perturbation independent} and typically equals the Lyapunov exponent of the underlying classical dynamics.
More recently a completely classical interpretation of this so-called Lyapunov decay has been given 
\cite{veble} 
in terms of {\em the classical Loschmidt echo} and a theory for exponents $\lambda$ has been developed in terms of the full spectrum of Lyapunov exponents for classical dynamical systems with few \cite{veble} or many \cite{veble2} degrees of freedom.

\subsubsection{Time dependent perturbation theory and linear response approximation.}
In the opposite regime, where the scaled Planck constant is bigger than the perturbation strength $\delta$,
one can use time-dependent quantum perturbation theory to second order to derive a simple
linear response formula for fidelity decay \cite{ProsenPRE02,fidelityreview}
\begin{equation}
F(t) = 1 - \frac{\delta^2}{\hbar^2}\int_0^t \dd t' \int_0^t \dd t'' C(t',t'') + {\cal O}((\delta/\hbar)^4)
\end{equation}
in terms of 2-point time correlation function of the perturbation
\begin{equation}
C(t',t'') = \ave{\tilde{V}(t')\tilde{V}(t'')} - \ave{\tilde{V}(t')}\ave{\tilde{V}(t'')}
\end{equation}
where $\tilde{V}(t) = U_0(-t) V U_0(t)$ is the perturbation operator in the interaction picture and 
$\ave{.} = \bra{\psi}.\ket{\psi}$
is an expectation value in the initial state.

From this formula - which can be viewed as a kind of Kubo-like linear response theory for dissipation of quantum
information, an interesting conclusion can be drawn: Fidelity decays faster for systems with 
slower decay of temporal correlations, or alternatively phrased, quantum system is more robust against
external perturbations if it relaxes to equilibrium faster.

One can actually go beyond the second order perturbation theory, and sum up the Born series for fidelity to all orders
in many specific situations \cite{fidelityreview}.
Since we are here mainly interested in qubit (spin 1/2) chains, we shall only review specific results for Kicked Ising 
chain \cite{ProsenPRE02}. We shall discuss three different specific values of system parameters, in all three we take
$J=1.0, h_\z = 1.4$:
(a) Integrable regime of transverse field $h_\x = 0$, (b) weakly non-integrable regime with 
$h_\x = 0.4$, and (c) strongly non-integrable regime with $h_\x = 1.4$.
All the results, correlation functions and fidelity, are for random initial states, which can be interpreted as pure
states of maximum quantum information, and averaging over random states is equivalent to a tracial state
$\ave{.} = 2^{-L}\tr(.)$.
We consider the evolution operator (\ref{eq:floquet}) and perturb it by changing the magnetic field 
such that the
perturbation is generated by the transverse component of the magnetization $M=\sum_j \s{\z}{j}$, namely
$U_\delta(t) = [U(J,h_\x,h_\z) \exp(-\ii \delta M)]^t$.

 \begin{figure}
       \quad\quad\quad\quad\quad\quad\includegraphics[width=14.0cm,angle=-90]{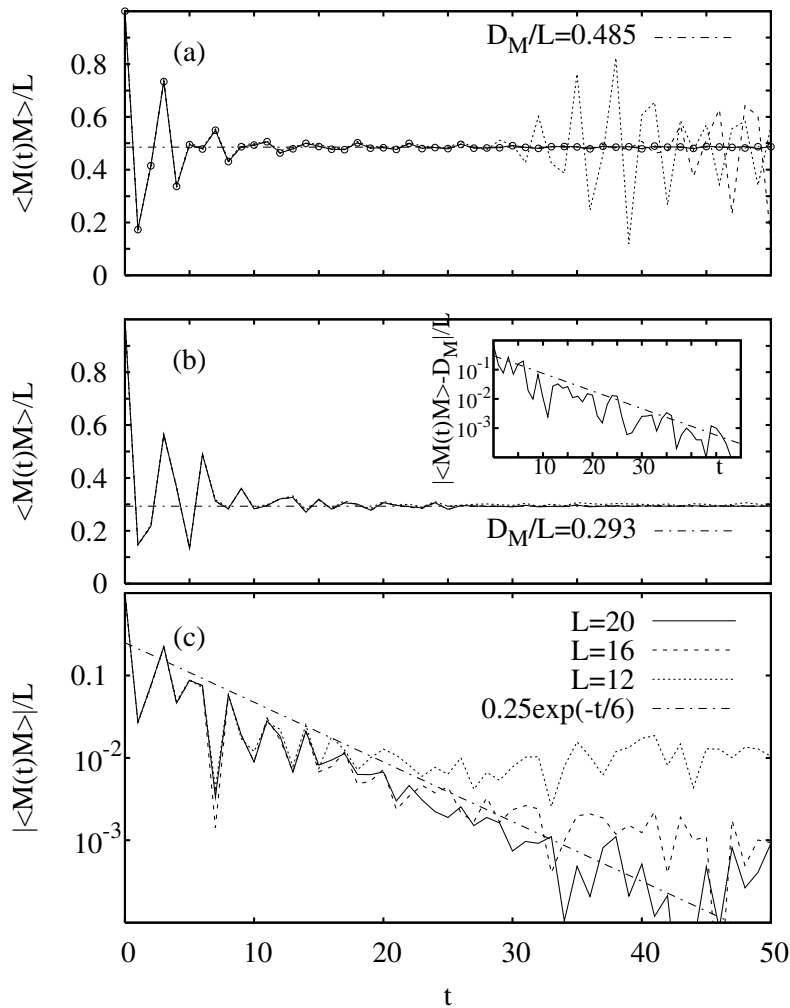}
       \caption{ Correlation decay for three cases of finite KI model: (a) integrable $h_{\rm x}=0$, (b)
       intermediate $h_{\rm x}= 0.4$, and (c) ergodic $h_{\rm x}=1.4 $, whereas $J=1.0,h_{\rm z}=1.4$, and
       for different sizes $L=20,16,12$ [solid-dotted connected curves, almost indistinguishable in (a,b)].
       Circles in (a) show exact result for $L=\infty$.
       }
       \label{fig:corr}
\end{figure}

\begin{figure}
       \quad\quad\quad\quad\quad\quad\includegraphics[width=14.0cm,angle=-90]{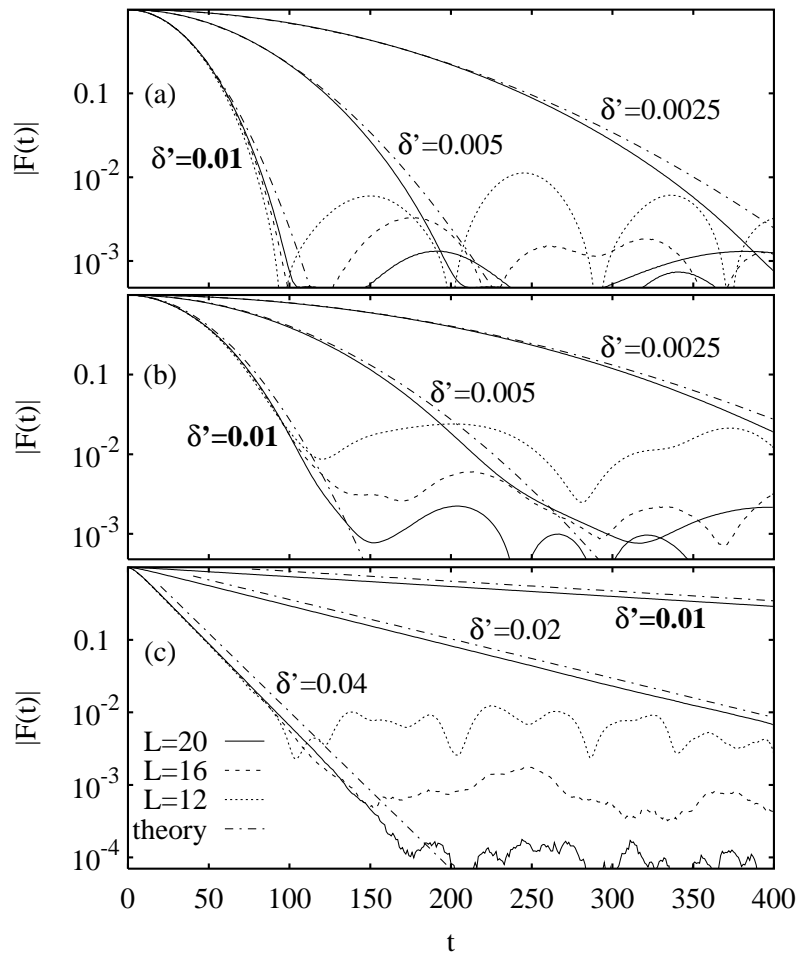}
       \caption{Average fidelity amplitude (absolute value of it) for three cases of finite KI: (a) integrable $h_{\rm x}=0$, (b)
       intermediate $h_{\rm x}= 0.4$, and (c) ergodic $h_{\rm x}=1.4 $, whereas $J=1.0,h_{\rm z}=1.4$, and
       for different sizes $L=20,16,12$, and different scaled perturbations $\delta'$. Chain curves give theoretical
       predictions \cite{ProsenPRE02,fidelityreview}.
       }
       \label{fig:fid}
\end{figure}

In the integrable case,  or in general, in non-ergodic case, where the correlation function 
$C(t',t'')=\ave{M(t')M(t'')}$ (note that $\ave{M(t)}\equiv 0$) approaches
a non-vanishing plateau value as $|t'-t''| \to \infty$, namely time averaged fluctuation of magnetization 
$ D_M = \lim_{t\to\infty}(1/t^2) \int_0^t \dd t' \int_0^t  \dd t'' C(t',t'') $ is non-vanishing $D_M\neq 0$,
one can sum up Born series to all orders, yielding a Gaussian decay of
fidelity
\begin{equation}
F(t) = \exp(-(t/\tau_{\rm ne})^2),\quad \tau_{\rm ne} = D^{-1/2}\delta^{-1}.
\label{eq:gauss}
\end{equation}
The only assumptions are that $t$ is long enough for the time average in the definition of $D_M$ to converge, and short enough that fidelity is still above the finite size plateau value $F^* \sim 1/2^L$. Note that $D_M$ can be considered as an analog
of Drude weight or charge stiffness in the linear response solid state transport theory (see e.g. \cite{Zotos}).
In figs.(\ref{fig:corr},\ref{fig:fid}) (a,b) one can observe the correlation plateaus of correlation functions 
for integrable and weakly non-integrable cases and respective good agreement with a Gaussian decay of fidelity (\ref{eq:gauss}). Note that in the integrable case the correlation function and the plateau $D_M$ have been calculated
also analytically \cite{Prosen00}. Note that for comparing different lattice sizes $L$ a size-scaled value of the 
perturbation strength $\delta' = \delta \sqrt{L/L_0}$, with $L_0=24$, 
has been fixed rather than $\delta$ itself.

One the other hand, for sufficiently strong integrability breaking, say in case (c) of KI model,
the correlation function $C(t'-t'')=C(t',t'')$ decays to zero in TL, which can be interpreted as quantum mixing behviour. It has been found that quantum mixing behaviour typically corresponds to random matrix (quasi)energy level statistics, see e.g. Ref.\cite{ProsenPRE99} and the 
subsequent sections of the present paper. We shall call such behaviour the regime of quantum chaos.
Here again, Born series for fidelity can be summed up to all orders, yielding an exponential
decay \cite{ProsenPRE02}
\begin{equation}
F(t) = \exp(-t/\tau_{\rm m}),\quad \tau_{\rm m} = 1/(2 \sigma \delta^2)
\label{eq:ne}
\end{equation}
where $\sigma=\int_0^t C(t')\dd t'$ is a transport coefficient.
The assumptions for validity of (\ref{eq:ne}) 
are that $t$ is larger than a characteristic mixing time scale on which $C(t)$ decays
and short enough so that finite size effects in quantum correlation function $C(t)$ does not yet affect
the transport coefficient $\sigma$.
This regime of fidelity decay is sometimes referred to as the Fermi Golden Rule regime.
Again, as demonstrated in figs.~(\ref{fig:corr},\ref{fig:fid})c the agreement of numerical data for KI model
with the theory is excellent.

Note that decay time of fidelity scales as $\propto \delta^{-2}$ for ergodic and mixing dynamics, and
as $\propto\delta^{-1}$ for non-ergodic dynamics, making the latter much more sensitive to small perturbations.

\subsubsection{Random matrix theory (RMT).}
Complex quantum systems can often be well described by statistical ensembles of Hamiltonians \cite{rmt}.
Assuming that $H_0$ and $V$ both belong to canonical ensembles of Gaussian random matrices, one
can evaluate ensemble averages of the fidelity and relate $C(t)$ to the spectral form factor of the 
underlying random matrix ensemble. 
The perturbative (linear response) theory has been developed in Ref.~\cite{GPS04}, see also 
Ref.~\cite{tomsovic,frahm}. Furthermore a non-perturbative (super-symmetric) averaging has been
successfully applied to obtain exact expressions for average fidelity amplitude in the most interesting cases 
\cite{stoeckmann}. RMT theory of fidelity has been successfully applied to chaotic quantum systems and even
to several experimental situations \cite{experiments}.

Following more applied philosophy, groups from Toulouse and Como have performed a series
of numerical experiments \cite{dima} analyzing the robustness of several reasonable models of
quantum computer  hardware under small imperfections, being either due to (static) unwanted
inter-qubit coupling or due to stochastic (noisy) unwanted coupling to the environment, when
performing quantum algorithms simulating various toy models of classical and quantum single-particle
chaos \cite{frahm}, like for example quantum kicked rotator \cite{dimakr}.
The numerical results are in line with a general linear response theory, stating that static perturbations are
in general more dangerous than noisy ones. 

\subsection{Decoherence and entanglement between weakly coupled systems}

Similar thinking as in the previous subsection can be applied to perhaps even more fundamental
problem of quantum physics, to the problem of decoherence \cite{zurekreview}.
Here we shall limit ourselves to an abstract unitary model of decoherence, where we treat a complete
unitary evolution over two subsystems, a central system C, and an environment E, and then address
a relevant information about the central system (i.e. the part of the system which is
of physical interest) by partially tracing over the environment.

Such a discussion can indeed be followed with a close link to the problem of Loschmidt echoes, by writing
the total Hamiltonian $H_\delta = H_0 + \delta V$ as an ideal (unperturbed) separable evolution
$H_0 = H_{\rm C}\otimes \one_{\rm E} +  \one_{\rm C}\otimes H_{\rm E}$ perturbed by a small coupling $V$
between the system and the environment.
We shall also assume that we start from initial pure state which is a product state 
$\ket{\psi} = \ket{\psi_{\rm C}}\otimes\ket{\psi_{\rm E}}.$ We are interested in the properties of a generally
mixed state of the central system obtained by partial tracing over the environment
\begin{equation}
\rho_{\rm C}(t) = \tr_{\rm E}\left[ U_\delta(-t) \ket{\psi}\bra{\psi} U_\delta(t) \right]
\end{equation}
Then, under an ideal evolution, the state of the system remains a product state at all times, so the state of
the central system remains pure and there is no decoherence. Decoherence is usually characterized
in terms of decaying off-diagonal matrix elements of $\rho_{\rm C}$ in a suitable pointer state basis, for example 
in the eigenbasis of the central system Hamiltonian $H_{\rm E}$. In fact for certain special forms of the
perturbation operator $V$, the magnitudes of off-diagonal matrix elements of $\rho_{\rm C}$ can be literally written
as fidelities, or quantum Loschmidt echoes, in the evolution of the environment perturbed by system-environment coupling \cite{strunz}.
In such framework, the Lyapunov regime of fidelity decay corresponds exactly to the Lyapunov growth of
decoherence discussed by Zurek and Paz \cite{zurek}.

However, another interesting indicator of decoherence is the growth of {\em bi-partite entanglement}
between the system and the environment. Perhaps this notion is even more general since it does not
depend on a particular choice of the pointer state basis.
Since the state of the universe (central system + environment) is always pure, the
characterization of entanglement is easy, say in terms of Von Neuman entropy of $\rho_{\rm C}$,
$S(t) = -\tr_{\rm C} \rho_{\rm C} \log \rho_{\rm C}$, or linear entropy
$S_2(t) = -\log \tr_{\rm C} \rho_{\rm C}^2(t)$ which is a negative logarithm of purity $P(t) = \tr \rho^2_{\rm C}(t)$.
Usually, the quantities $S(t)$ and $S_2(t)$ essentially give equivalent results - namely they can 
both be interpreted as the logarithm of an effective rank of the
state $\rho_{\rm C}$,  but the linear entropy, or purity, is more amenable to analytical calculations.

Again considering chaotic models for the central system and the environment, Miller and Sarkar \cite{miller} (see also \cite{bandy})
have been able to observe the 'Lyapunov regime' of entanglement growth, namely $S(t)$ is for sufficiently
strong perturbations $\delta$ found to grow linearly $S(t) = h t$ with the rate $h$ which is perturbation independent and
given by (the sum of positive) classical Lyapunov exponents of the unperturbed, separated (sub)systems.

On the other hand, in the purely quantum regime of small perturbation $\delta$, again typically smaller than
effective Planck constant, one can use time-dependent perturbation theory and derive perturbation dependent entanglement entropy 
\cite{znidaric,tanaka}, namely the purity can be explicitly expressed as 
$P(t) \sim 1 - \delta^2 {\cal C}(t)$ where ${\cal C}(t)$ is a particular integrated correlation function of the perturbation.
In this regime we again find that quantum systems, and quantum environments, which display faster
relaxation to equilibrium, are more robust against decoherence due to typical couplings.
For studies of bi-partite entanglement, in particular in KI model system, see Refs.~\cite{ps02,pineda}.

There exists even closer connection to fidelity theory, namely one can prove a general inequality 
\cite{znidaric,PSZ} stating that purity is always bounded by the square of fidelity
\begin{equation}
(F(t))^2 \le P(t). 
\end{equation}
In other words: $\log(1/F(t)^2)$ gives an upper bound for the growth of the linear entanglement entropy,
$S_2(t) \le -2 \log F(t)$.

In a slightly different context, {\em bit-wise} entanglement between a pair of qubits of a quantum register representing a 
time dependent quantum state, where the rest of the register is considered as an environment, 
has been demonstrated to be an indicator of quantum chaos \cite{como} and even a signature of classical chaos \cite{hu}.

\section{Efficiency of classical simulations of quantum systems}

\label{sect:MPS}

In the theory of classical dynamical systems there is a fundamental difference between integrable and chaotic systems as outlined in section \ref{sect:definitions}. 
Chaotic systems, having positive algorithmic complexity, unlike the 
integrable ones, cannot be simulated for arbitrary times with a finite amount of information about 
their initial states. Computational complexity of individual chaotic trajectories is linear in time, however, if one wants to describe statistical states (phase space distributions) or observables of chaotic classical systems, up to time $t$, exponential amount of computational resources is needed, 
typically $\propto \exp(h t)$ where $h$ is the Kolmogorov's dynamical entropy related to exponential sensitivity to initial conditions.  But on the other hand, how difficult is it to simulate isolated and bounded quantum systems of many interacting particles 
using classical resources?  In analogy with the classical (chaotic) case, we might expect that the best classical simulation of typical quantum systems (in TL) is exponentially hard, i.e. the amount of computing resources is expected to grow exponentially with time.

Even though there is no exponential sensitivity to initial conditions in quantum mechanics,
there is a tensor-product structure of the many-body quantum state space which makes
its dimension to scale exponentially with the number of particles, as opposed to linear scaling in the
classical case, and due to presence of entanglement generic quantum time evolution cannot be reduced to (efficient) classical computation in terms of non-entangled 
(classical like) states. 
Here we propose the idea to use {\em the computation complexity of best possible 
classical simulation of quantum dynamics, as a measure of quantum algorithmic complexity}.
This section essentially reviews the article \cite{PZ07}.
We note that our proposal is different from existing proposals of quantum algorithmic
complexity \cite{gacs,vitanyi,berthiaume,mora}, namely we consider merely the classical
complexity of (best) classical simulations of quantum states.
In the sense of Mora and Briegel \cite{mora}, quantum algorithmic complexity {\em per unit time} 
of initially simple time-evolving quantum states propagated by locally interacting Hamiltonian $H$
is clearly vanishing, since an approximate quantum circuit which reproduces the state
after time $t$ is a simple repetition of Trotter-Suzuki decomposition of $\exp(-\ii H \delta t)$.

\subsection{Time dependent density matrix renormalization group:
How far can it go?}

Recently, a family of numerical methods for the simulation of interacting many-body systems
has been developed \cite{MPS} which is usually referred to as
time-dependent density-matrix-renormalization group (tDMRG), and which has been shown to
 provide an efficient classical simulation of some interacting quantum systems. 
Of course, it cannot be proven that tDMRG  provides the best classical simulation of quantum systems,
but it seems that it is by now the best method available. 
Simulations of locally interacting one-dimensional quantum lattices were actually shown rigorously to be efficient in the number $L$ of particles \cite{Osborn} (i.e., computation time and memory resources scale as polynomial functions of $L$ at fixed $t$, whereas here we are interested in the scaling of computation time and memory with physical time  $t$ (in TL, $L=\infty$), referred to as {\em time efficiency}.

In this section we address the question of time efficiency implementing tDMRG for a family of Ising Hamiltonian (\ref{eq:hamaut}) which undergoes a transition from integrable (transverse Ising) to non-integrable quantum chaotic regime as the magnetic field is varied. 
We mainly consider time evolution in operator spaces \cite{tdmrg}, say of density matrices of quantum states,
or (quasi) local operator algebras.  Note that time evolution of pure states is often ill defined in TL 
\cite{schumacherwerner}.  As a quantitative measure of time efficiency we define and compute the minimal dimension  $D_\epsilon(t)$ of matrix product operator (MPO) representation of quantum 
states/observables which describes time evolution up to time $t$ within fidelity $1-\mathcal{O}(\epsilon)$. Our main question concerns possible scaling of $D_\epsilon(t)$ for different types of dynamics, and
indeed we shall demonstrate a correspondence between, respectively, 
quantum chaos or integrability, and exponential or polynomial growth of $D_\epsilon(t)$.

The key idea of operator-valued tDMRG \cite{tdmrg} is to represent any
operator in a matrix product form, 
\begin{equation}
O_{\rm MPO} = \sum_{s_j} \tr{(A^{s_0}_0 A^{s_1}_1 \cdots A^{s_{L-1}}_{L-1})} 
\s{s_0}{0} \s{s_1}{1}\cdots \s{s_{L-1}}{L-1},
\label{eq:MPO}
 \end{equation}
in terms of $4 L$ matrices $A^{s_j}_j$ of
fixed dimension $D$. The number of parameters in the MPO representation
is $4LD^2$ and for sufficiently large $D$ it can
describe any operator. In fact, the minimal $D$ required equals to the 
maximal rank of the reduced super-density-matrix over all bi-partitions
of the chain. The advantage of MPO representation lies in the fact that
doing an elementary local one or two qubit unitary transformation 
$O'=U^\dagger O U $ can be done locally, affecting only a pair of neighboring
matrices $A^{s_j}_j$.
\par
\begin{figure}[h!]
\hfil\includegraphics[angle=-90,width=12.0cm]{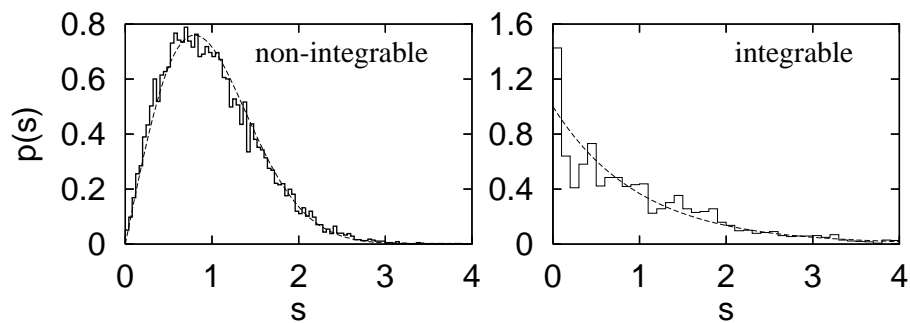}
\caption{Nearest neighbor LSD for $H_{\rm C}$ (left) and $H_{\rm R}$
(right) for $L=12$. Dashed curves are $p(s)=s \pi/2 \exp{(-\pi^2
s^2/4)}$ (left) and $p(s)=\exp{(-s)}$ (right), typical for chaotic and
regular systems, respectively\cite{Haake:00}. Eigenenergies $\in
[-9,9]$ were used and statistics for even and odd parity states were
combined.} \label{fig:lsd} 
\end{figure}

  
We will illustrate evolution of Ising chain (\ref{eq:hamaut}), with {\em open} boundary
conditions, for two different magnetic field
values: (i) an integrable (regular) case $H_{\rm R}=H'(1,0,2)$ with
transverse magnetic field, and (ii) non-integrable (quantum chaotic)
case $H_{\rm C}=H'(1,1,1)$ with$45^\circ$ tilted magnetic field.
To confirm that $H_{\rm C}$, and
$H_{\rm R}$, indeed represent generic quantum chaotic, and regular,
system, respectively, we calculated level spacing distribution (LSD) of
their spectra (shown in fig.~\ref{fig:lsd}). LSD is a standard
indicator of quantum chaos~\cite{Haake:00}. It displays characteristic
level repulsion for strongly non-integrable quantum systems, whereas
for integrable systems there is no repulsion due to existence of
conservation laws and quantum numbers. 
Evolution by tDMRG proceeds as described in \cite{MPS,tdmrg,PZ07} 
using an approximate Trotter-Suziki factorization, for some time step $\delta t$, 
of the evolution operator $U(t)=\exp(-\ii H t)$ in terms of  2-qubit gates. 
And for each two qubit gates, the matrices $A^{s_j}_j$ can
be updates using a singular value decomposition with some truncation error $\eta$. 
Interestingly, it has been found \cite{PZ07} that the sum
of all truncation errors up to time $t$, denoted by $\eta(t)$ (provided $\delta t$ 
is small enough so that the Trotter-Suzuki factorization error is negligible) is simply 
proportional to {\em fidelity error}, namely 
\begin{equation}
1-F(t) \approx c \eta(t)/\delta t,
\end{equation}
where
\begin{equation}
F(t)=\frac{|\tr{\{O_{\rm MPO}(t) O_{\rm
exact}(t)\}}|^2}{|\tr{\{O^2_{\rm MPO}(t)\}}||\tr{\{O^2_{\rm
exact}(t)\}}|},
\end{equation} 
and numerical constant $c \sim 1$ does not dependent on either $\delta t$, $D$ or $L$.
The central quantity we are going to study is $D_\epsilon (t)$ which is the minimal dimension $D$ of matrices $A^{s_i}_i$ in order for the total truncation error $\eta(t)$ to be less than some error tolerance $\epsilon$, for  evolution to time $t$.  In numerical experiments shown we take $\epsilon=10^{-4}$ 
except for simulating thermal states of quenched Hamiltonians where $\epsilon = 10^{-6}$.

\subsection{Simulating pure states}

\begin{figure}
       \centering
       \includegraphics[width=7.0cm,angle=-90]{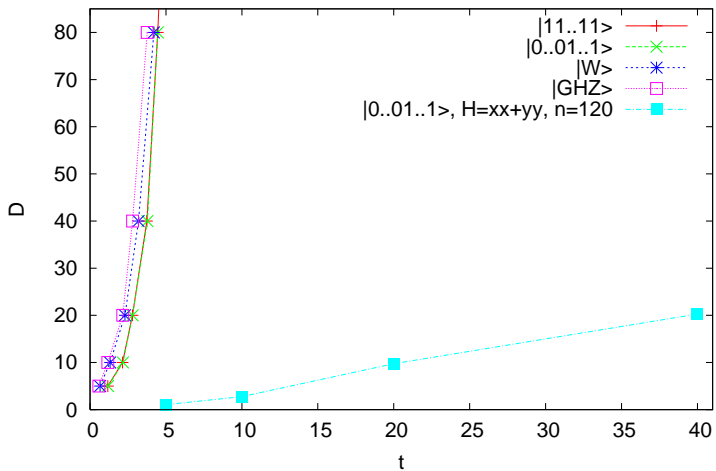}
       \caption{MPS rank $D_\epsilon(t)$ for simulating pure states with integrable transverse Ising model $H_{\rm R}$,
       except full squares which are for Heisenberg XX chain, starting from the initial states indicated in the legend
       (explanation: $\ket{W}=(\ket{10\ldots0} + \ket{01\ldots0}+\cdots_\ket{00\ldots 1})/\sqrt{L}$ and
         $\ket{GHZ}=(\ket{00\ldots 0}+\ket{11\ldots 1})/\sqrt{2}$).        
       Note that the full squares corresponds to the same example as studied in Ref.\cite{SchWhite}.
       }
       \label{fig:pure}
\end{figure}

For a reference we start by investigating the evolution of pure states following the basic tDMRG algorithm \cite{MPS}.
We studied time efficiency of simulation of {\em pure states} in 
Schr\" odinger picture, for which many examples of efficient applications exist, however all for initial states of rather particular structure, typically corresponding to low energy sectors of few quasi-particle excitations or to low dimensional invariant subspaces. 
Treating other, typical states, e.g. eigenstates of unrelated Hamiltonians, linear combinations of highly excited states,  or states chosen randomly in the many-particle
Hilbert space, we found that, irrespectively of integrability of dynamics,
 tDMRG is {\em not} time-efficient, i.e. $D_\epsilon(t)$ typically grows exponentially in time even in the integrable case of transverse field (consistently with a linear growth of entanglement entropy~\cite{Calabrese:05}). Numerical results are summarized in fig.~\ref{fig:pure}.

%We find that, unless the initial state is a (i) liner combination of low energy
%quasi-particle excitations (for the integrable case), or (ii) for some reason belongs
%to a low dimensional invariant subspace, generic behavior is $D_{\epsilon}(t) = \exp( {\rm const} t)$

\subsection{Simulating local observables}

\begin{figure}[h!]
\hfil\includegraphics[width=7.0cm,angle=-90]{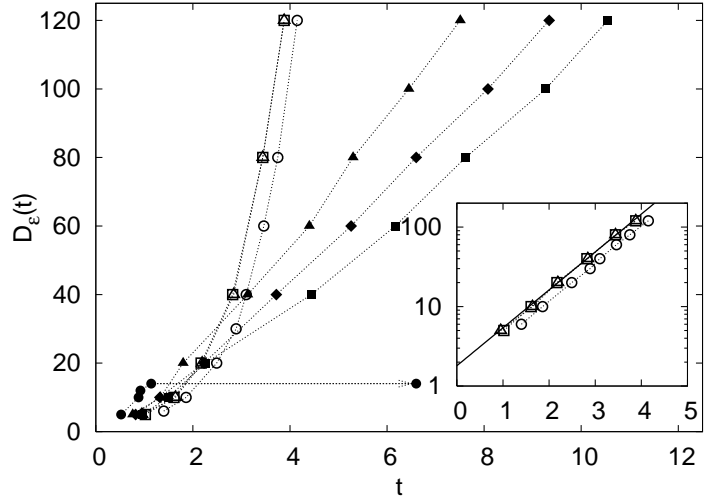}
  \caption{$D_\epsilon(t)$ for local initial operators.
  We consider three cases $O(0)=\sigma^{\rm x,y,z}_{L/2}$ (empty circles, squares and triangles), for non-integrable evolution $H_{\rm C}$,  and four cases, $O(0)=\sigma^{\rm x,y}_{L/2}$ (full squares, diamonds), $\sigma^{\rm z}_{L/2-1}\sigma^{\rm y}_{L/2}$ (full triangles), all with infinite index, 
  and $O(0) = \sigma^{\rm z}_{L/2-1}\sigma^{\rm z}_{L/2}$ (full circles) with index 2, for integrable evolution $H_{\rm R}$. In the inset we plot the data for the non-integrable case $H_{\rm C}$ 
  in semi-logarithmic scale, and the
   full line in the inset illustrates exponential growth $\propto e^{1.1 t}$. Full squares and diamonds are for $L=40$, otherwise $L=20$. 
 }
\label{fig:local}
\end{figure}

We continue by discussing the time efficiency of operator-valued tDMRG method using MPOs (\ref{eq:MPO}).
Let us first study the case where the initial operator is a local operator in the center of the lattice
$O(0) = \sigma^s_{L/2}, s\in\{\rm x,y,z\}$. In the integrable case time evolution $O(t)$ can be computed 
exactly in terms of Jordan-Wigner transformation and Toeplitz determinants \cite{jacoby}, 
however for initial operators with infinite {\em index}\footnote{
 Index of a product operator $O$ [Sect.2, 1st of Refs.\cite{jacoby}] is half the number
of fermi operators in Jordan-Wigner transformation of $O$ and is a conserved quantity for $H_{\rm R}$.}
like e.g for $\sigma^{\rm x,y}_{L/2}$,  $L\to\infty$, the evolution is rather complex and the effective number of terms (Pauli group elements) needed to span $O(t)$ grows exponentially in $t$.
In spite of that, our numerical simulations shown in fig.~\ref{fig:local} strongly suggest the
linear growth $D_\epsilon(t) \sim t$ for initial operators with infinite index.
Quite interestingly, for initial operators with {\em finite} index, $D_{\epsilon}(t)$ saturates to
a finite value, for example $D_{\epsilon}(\infty)=4$ for $\sigma_{L/2}^{\rm z}$,  or
$D_{\epsilon}(\infty)=16$ for $\sigma_{L/2-1}^{\rm z} \sigma_{L/2}^{\rm z}$.
In non-integrable cases the rank has been found to grow exponentially, $D_\epsilon(t) 
\sim \exp(h_{\rm q} t)$ with exponent $h_{\rm q}$ which does {\em not} depend on
$\epsilon$, properties of $O(0)$ or $L$, for large $L$. For $H=H_{\rm C}$ we find $h_{\rm q} = 1.10$.

\subsection{Simulating extensive observables}

\begin{figure}[h!]
\hfil\includegraphics[width=7.0cm,angle=-90]{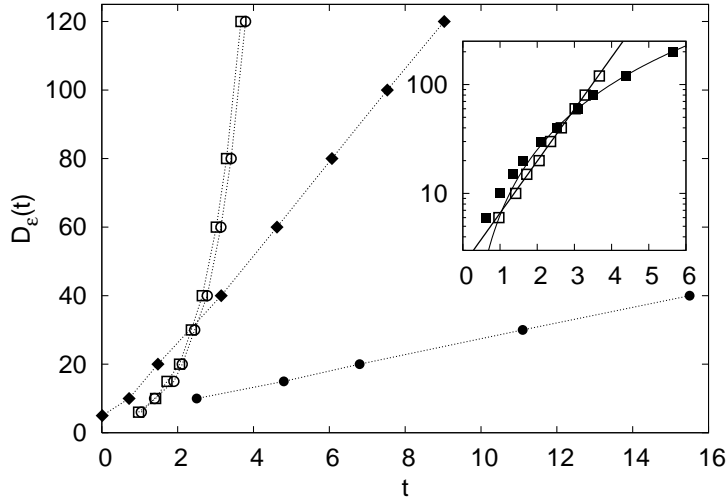}
  \caption{$D_\epsilon(t)$ for extensive initial operators. For both Hamiltonians 
  $H_{\rm C}$, $H_{\rm R}$ we take $O(0)=\sum_j{\sigma^{\rm x}_j}$ (empty, full squares) with
  infinite index, and $O(0)=H'(1,0,1)$ (empty, full circles) with index 1. For $H_{\rm R}$ we also show the case $O(0)=\sum_j{\sigma^{\rm z}_j \sigma^{\rm z}_{j+1}+\sigma^{\rm y}_j \sigma^{\rm y}_{j+1}}$ (full diamonds) with index 1 and 2.  In the semi-log inset we illustrate exponential increase $\propto e^{1.1 t}$ (full straight line) for $H_{\rm C}$ and polynomial $\sim t^2$ (full curve) for $H_{\rm R}$. For full circles $L=64$, otherwise $L=32$.}
\label{fig:transl}
\end{figure}

In physics it is often useful to consider extensive observables, for instance
translational sums of local operators, e.g. the Hamiltonian $H$ or the total magnetization 
$M^s=\sum_{j=0}^{L-1} \sigma_j^s$. As opposed to local operators, extensive initial operators,
interpreted as W-like states in operator space, contain some long-range `entanglement'
so one may expect that tDMRG should be somewhat less efficient than for local operators.
Indeed,  in the integrable case we find for extensive operators with 
{\em finite} index that  $D_\epsilon(t)$ 
does no longer saturate but now grows linearly, 
$D_\epsilon(t) \sim t$, whereas for extensive operators with {\em infinite} index
the growth may be even somewhat faster, most likely quadratic
$D_\epsilon(t) \sim t^2$ but clearly slower than exponential. 
In the non-integrable case, we again find exponential growth
$D_\epsilon(t) \sim \exp(h_{\rm q} t)$ with the same exponent $h_{\rm q}=1.10$ 
as for local initial observables.
The results are summarized in fig.~\ref{fig:transl}. 

\subsection{Simulating thermal states}

\begin{figure}[ht]
\hfil\includegraphics[width=7.0cm,angle=-90]{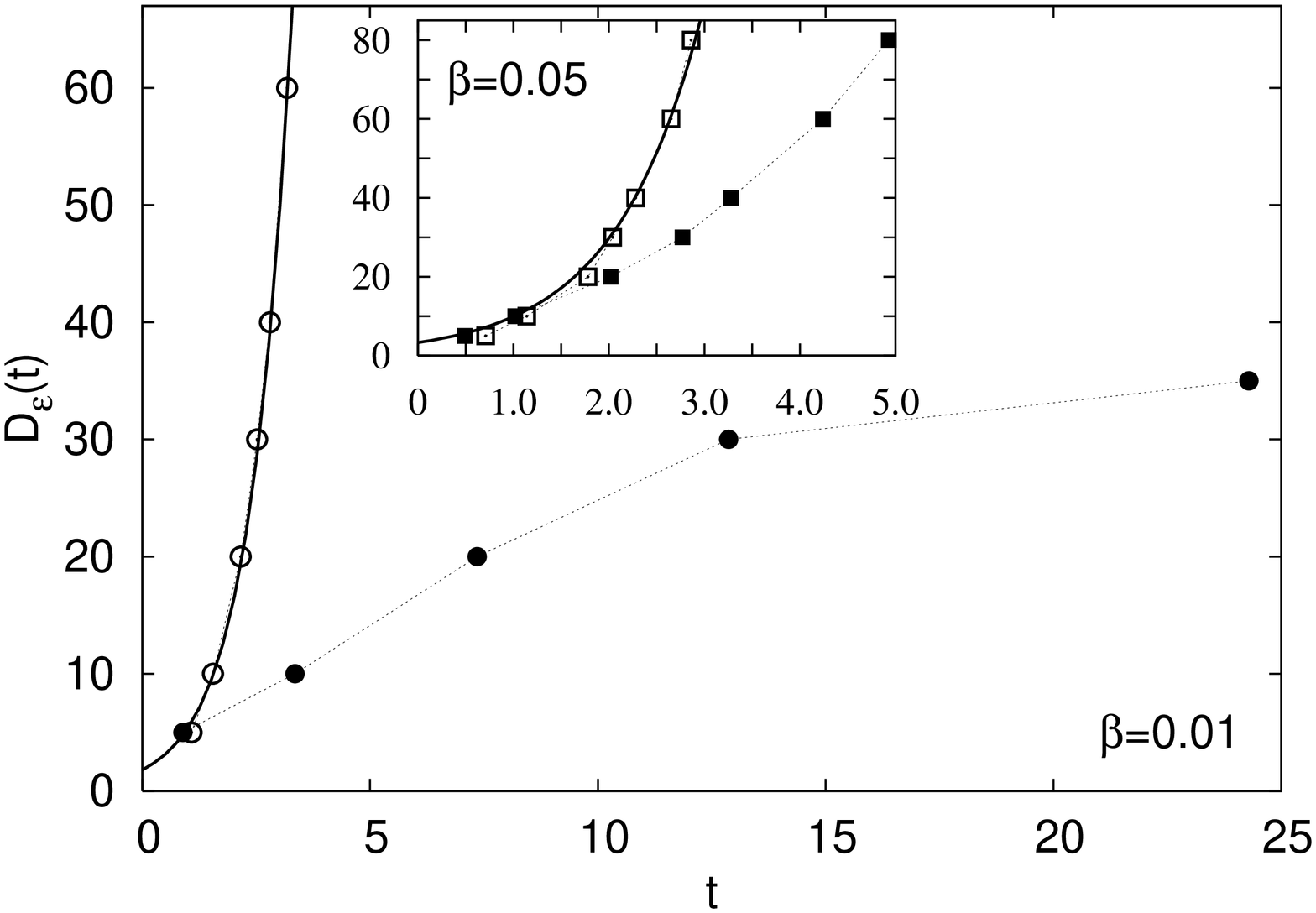}
  \caption{
  $D_\epsilon(t)$ for thermal states of $H_0$ with $\beta=0.01$ ($\beta=0.05$ in inset),
  for evolution with $H_{\rm C}$ (open symbols) and $H_{\rm R}$ (full symbols) at $L=40$.
  Solid curves again indicate exponential increase $\propto e^{1.1 t}$.}
\label{fig:rhoT}
\end{figure}
In the last set of numerical experiments we consider time efficiency of the evolution of a thermal initial state
$O(t) = Z^{-1} \exp(-\beta H_0)$ under a sudden change of the Hamiltonian at $t=0$,
namely $H(t < 0) = H_0=H'(1,0,1), H(t > 0) = H_1$. Again, we treat two situations: in the first case we consider change after which the Hamiltonian remains integrable, $H_1 = H'(1,0,2) = H_{\rm R}$, while in the other case the
change breaks integrability of the Hamiltonian, $H_1 = H'(1,1,1) = H_{\rm C}$.
The initial state is prepared by means of imaginary time
tDMRG from initial identity operator using the same MPO rank $D$ as it is later
used for real time dynamics.
We find, consistently with previous results, that at high temperature ($\beta \ll 1$) the
rank $D_\epsilon(t)$ grows very slowly,
perhaps slower than linear, in the integrable case, 
and exponentially $D_\epsilon(t) \sim \exp(h_{\rm q} t)$, in the non-integrable
case. Interestingly, at lower temperatures we find exponential growth in both cases, even in the integrable one. This is not unreasonable as the initial (thermal) state can be
expanded in a power series in $\beta$ and the higher orders $H_0^p$ become less local
with longer entanglement range as we increase the power $p$.
These results are summarized in fig.~\ref{fig:rhoT}. 

\section{Quantum chaos and far from equilibrium quantum transport}

\label{sect:transport}

In this section we would like to demonstrate the connection between
quantum chaos, (non)integrability and transport in non-equilibrium steady
states of interacting quantum chains \cite{mejia05}.
Within the linear response theory, the property of quantum mixing (which is
typically implied by quantum chaos, i.e. by the validity of random matrix level statistics
\cite{ProsenPRE99}, see also 
correlation functions in sect.\ref{sect:losch}), is typically synonymous for
normal quantum transport since it implies {\em finite} Kubo transport coefficients \cite{ kubo}
(provided temporal correlation functions decay fast enough).

However, here we would like to address the connection between quantum chaos and transport in
far-from-equilibrium steady state, which may be beyond applicability of linear
response. In particular, we are interested in the validity \cite{lebowitz} of the
Fourier  law $J=-\kappa\nabla T$ in quantum chains, 
relating the  macroscopic heat flux
$J$ to the temperature gradient $\nabla T$. 

To  investigate  this problem  one  has  to deal  with  a  finite open  system
connected to  heat baths. Here  we consider an interacting  quantum spin-$1/2$
chain (\ref{eq:hamaut}) which exhibits  the transition from integrability to  quantum chaos as a
parameter, e.g.  the magnetic field, is varied. The standard treatment of this
problem  is based  on the  master  equation, thus  limiting numerical investigations  to
relatively small system  sizes. By using this method,  in an interesting paper
\cite{saito},  the  decay  of  current  correlation function  in  a  model  of
non-integrable chain of quantum spins  is computed. However, these results were
not  fully conclusive  and the  conclusions rely  on linear  response  theory. 
Also, in \cite{hartmann} Lindblad formalism  was used to study the validity of
Fourier's law for  different type of spin-spin interaction. 

Here we describe a different approach (see Ref.\cite{mejia05} for details) 
namely we follow the evolution of the system described by a  {\em pure} state, 
which is {\em  stochastically} coupled to an idealized model of  heat baths. 
Stochastic coupling is realized  in terms of a
local measurement  at the  boundary of the  system and {\em stochastic}  but {\em unitary}
exchange of  energy between the system and  the bath.  By this  method we have
been  able to  perform very  effective  numerical simulations  which allow  to
observe a  clear energy/temperature  profile and to  measure the  heat current
$J$.  Again we consider Ising spin chain in the magnetic field (\ref{eq:hamaut})
of size $L$, where the first and the last spin are coupled to thermal baths at temperatures
$T_{\rm l}$ and $T_{\rm r}$, respectively.
In the  non-integrable regime where the spectral  statistics is described
by RMT - the regime  of quantum chaos -  we found
very accurate Fourier  law scaling $J/\Delta T \propto 1/L$,  where $L$ is the
size of the  chain. In the integrable and near-integrable ({\em non-ergodic}) regimes instead, we
found that the heat transport is ballistic $J \propto L^0$.

Let us describe the numerical simulations. 
Again we consider the autonomous model (\ref{eq:hamaut}) at three 
particular cases: ({\it i}) {\em
  quantum chaotic case} $H=H'(-2,2,3.375)$ at  which LSD agrees
with RMT and thus corresponds to  the regime of quantum chaos, ({\it ii}) {\em
  integrable  case} $H=H'(-2,0,3.375)$,  at which  LSD is  close  to the
Poisson    distribution,   and   ({\it    iii})   {\em    intermediate   case}
$H=H'(-2,2,7.875)$ at which the  distribution LSD shows and intermediate
character of weak level repulsion and exponential tail \cite{mejia05}. 

Let us now turn  to study the energy transport in this model  system.  To this end
we need  to couple both ends  of the chain  of spins to thermal  baths at
temperatures  $T_{\rm r},T_{\rm l}$.  We have  devised a  simple way  to simulate  this coupling,
namely  the state  of  the spin  in  contact with  the  bath is  statistically
determined by a  Boltzmann distribution with parameter $T$.  Our model for the
baths is analogous to the stochastic thermal baths used in classical
simulations \cite{larralde}  and we  thus  call it  a \emph{quantum  stochastic thermal
bath}.  In the representation basis of $\s{\z}{n}$  the wave function at time $t$ can
be written as
\begin{equation}
\ket{\psi(t)}=\!\!\!\!\sum_{s_0,s_1,\ldots,s_{L-1}}\!\!\!\!\!\! C_{s_0,s_1,\ldots,s_{L-1}}(t)
\ket{s_0,s_1\ldots s_{L-1}}\ ,
\end{equation}
where $s_n  =0,1$ represents  the \emph{up}, \emph{down}  state of  the $n$-th
spin,  respectively.  The  wave  function at  time  $t$ is  obtained from  the
unitary  evolution  operator   $U(t)  =  \exp(-\ii H t)$.   The
interaction  with the bath is  not included  in the  unitary  evolution. 
Instead, we  assume that  the spin  chain and the bath interact  only at
discrete times with period $\tau$ at  which the states of the leftmost ($s_0$)
and  the  rightmost ($s_{L-1}$)  spins  are  stochastically  reset. Thus,  the
evolution  of  the  wave function  from  time  $t$  to  time $t+\tau$  can  be
represented as
\begin{equation}
\ket{\psi(t+\tau)} = \Xi(\beta_{\rm l},\beta_{\rm r}) U(\tau)\ket{\psi(t)} \ ,
\end{equation}
where  $\Xi(\beta_{\rm l},\beta_{\rm  r})$ represents  the  {\em stochastic}
action of the  interaction with the left and  right baths at temperatures
$\beta_{\rm l}^{-1}$ and $\beta_{\rm r}^{-1}$ respectively.

The action of $\Xi(\beta_{\rm l},\beta_{\rm r})$ takes place in several steps:
\begin{itemize}
\item ({\it  i})  The  wave  function  is  first  rotated  by  the  angle
\mbox{$\alpha  =  \arctan(h_\x/h_\z)$} to  the  eigenbasis  of the  components
$\sigma_\mathrm{l} = \vec{h}\cdot\vec{\sigma}_0/h$, $\sigma_\mathrm{r}
=  \vec{h}\cdot\vec{\sigma}_{L-1}/h$ of  the edge  spins along  the  field $\vec{h}=(h_\x,0,h_\z)$, 
that is
$\ket{\psi}  \rightarrow e^{-\ii\alpha(\sigma_0^\y+\sigma_{L-1}^\y)/2}\ket{\psi}$.
Here, $h=|\vec{h}|$ stands for the magnitude of the magnetic field.

\item   ({\it    ii})   A    local   measurement   of    the   observables
$\sigma_\mathrm{l}$,  $\sigma_\mathrm{r}$ is performed.  Then the 
state  of the
spins  at  the  borders   collapses  to  a  state  ($s_0^*$,$s_{L-1}^*$)  with
probability
\begin{equation}
p(s_0^*,s_{L-1}^*)=\sum_{s_1,\ldots,s_{L-2}}
|C_{s_0^*,s_1,\ldots,s_{L-2},s_{L-1}^*}|^2 \ .
\end{equation}
So, after choosing $s_0^*,s_{L-1}^*$,   we    put   all    coefficients
$C_{s_0,s_1,\ldots,s_{L-1}}$  with $(s_0,s_{L-1})  \neq  (s_0^*,s_{L-1}^*)$ to
zero.

\item  ({\it iii})  The new  state of  the edge  spins  $(s_0,s_{L-1})$ is
stochastically  chosen. After  this action  simulating the  thermal interaction
with the baths each of the edge spins is set to  {\em down}, ({\em up})
state with probability  $\mu$,($1-\mu$).  If the new state of $s_0$, or $s_{L-1}$,
is different than the one after step (ii), then a simple unitary spin flip
is performed to the wave-function.
The probability $\mu(\beta)$ depends
on the canonical temperature of each of the thermal baths:
\begin{equation} \label{eq:canonical}
\mu(\beta_j)  =   \frac{e^{\beta_jh}}{e^{-\beta_jh}  +  e^{\beta_jh}}
\quad ; \quad j \in \{\mathrm{l,r}\} \ .
\end{equation}

\item  ({\it  iv}) Finally,  the  wave function  is  rotated  back to
the         $\sigma^z_n$         basis,        $\ket{\psi}         \rightarrow
e^{\ii\alpha(\sigma_0^y+\sigma_{L-1}^y)/2}\ket{\psi}$.
\end{itemize}

This completes the description of  the interaction with the quantum stochastic
bath.   This interaction  thus (periodically)  resets the  value of  the local
energy $h\sigma_{\rm l,r}$  of the spins in contact  with the baths.  
Therefore, the  value of $\tau$
controls the strength of the coupling to  the bath. We have found that, in our
units, $\tau=1$  provides an optimal  choice.  We have  nevertheless performed
simulations for other values of  $\tau$ with qualitatively similar results. In
particular, for weak couplings ($\tau\gtrsim1$) the heat conductivity does not
depend on the coupling strength .

Note  that our  method does  not correspond  to a  standard stochastic unraveling  of a
master equation for the  density operator (\emph{e.g.}, in  the Lindblad
form) \cite{lindblad}. 
However,  we have tested  that, using our  prescription, averages
over  the ensemble  of  quantum trajectories  or  time averages  of one  given
quantum  trajectory are  sufficient to  reconstruct a  density matrix
operator $\rho = \overline{\ket{\psi(t)}\bra{\psi(t)}}$ that correctly describes the  internal thermal state of the system in
and out of equilibrium. 
For each run  the initial wave-function $\ket{\psi(0)}$ of the
system is  chosen at random.  The  system is then evolved  for some relaxation
time $\tau_{\rm rel}$  after which it is assumed to  fluctuate around a unique
steady  state.  Measurements  are  then  performed as  time  averages  of  the
expectation  values   of  suitable  observables.   We   further  average  these
quantities over different random realizations of ``quantum trajectories''.
\begin{figure}[!t]
\begin{center}
\includegraphics[scale=0.6]{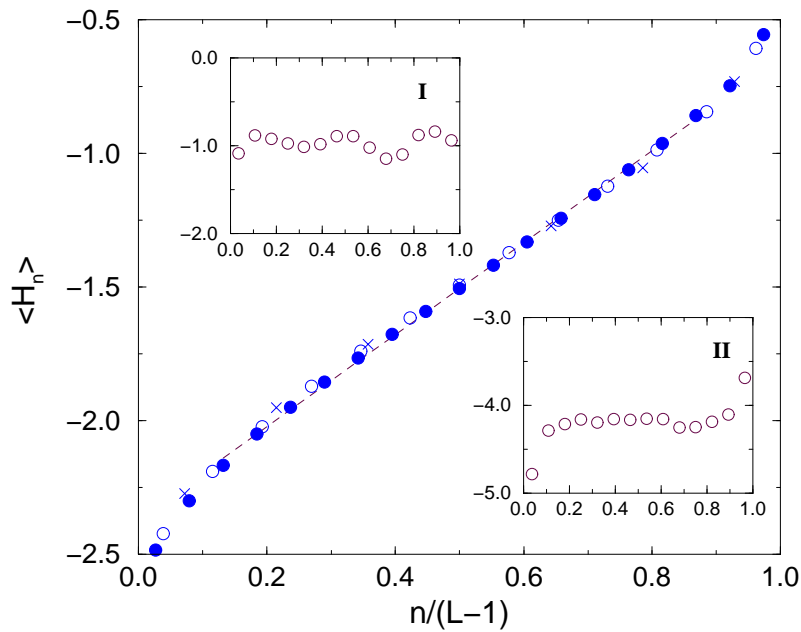}
\caption{\label{fig:3}
  Out of  equilibrium energy profile  $\ave{H_n}$ for the chaotic  chain.  The
  temperatures of the baths $T_{\rm l} =  5$ and $T_{\rm r} = 50$, are both in
  the high  temperature regime.  Results for chains of size $L=8$ (crosses), 
  $L=14$ (open circles) and $L=20$
  (solid circles) are  shown.  The dashed line was obtained  from a linear fit
  of the data for $L=20$ for the $L-4$ central spins. Insets (I) and (II) show
  the energy  profile for the integrable and  intermediate cases respectively,
  for $L=15$.}
\end{center}
\end{figure}
In order  to compute the energy  profile we write the Hamiltonian (\ref{eq:hamaut}) in
terms of local energy density operators $H_n$:
\begin{equation} \label{eq:H_local}
H_{n}     =   J\sigma^{\rm x}_n\sigma^{\rm x}_{n+1}    +
\frac{\vec{h}}{2} \cdot \left(\vec{\sigma}_n  + \vec{\sigma}_{n+1}\right) \ .
\end{equation}
such that the total Hamiltonian (\ref{eq:hamaut}) can be written as $H=\sum_{n} H_n$
apart from the boundary corrections.

First we  have performed  equilibrium simulations in  order to show  that time
averaged  expectation values  of the  local energy  density can  be used  to determine 
canonical local temperature.  To this end we set the left and right
baths to the same temperature  $T$.  For low $T$, the energy per site $E = (1/L) \ave{H}$  saturates to a 
constant which, together with
the entire energy profile  $\ave{H_n}$, is determined by the  ground state.  However,
for larger $T > 1$, the  energy profile is constant within numerical accuracy,
and  numerical  simulations give  $E  \sim  -1/T$,  all results  being  almost
independent of  $L$ for $L\ge  6$. The numerical  data for $E(T)$ can  be well
approximated with a simple calculation  of energy density for a two-spin chain
($L=2$)  in a  canonical  state  at temperature  $T$,  namely
$E_{\rm  can}(T)=\tr  H_0   e^{-H_0/T}/\tr  e^{-H_0/T}$.   Therefore,  if  the
temperatures of both baths are in high $T$ regime, then we can define the
local  temperature  via  the  relation  $T  \propto  -1/E$.   We  stress  that
equilibrium numerical  data shown  are {\em insensitive} to  the nature  of dynamics
(consistent with results of Ref.\cite{jensen}), whether being chaotic, regular
or intermediate.
\begin{figure}[!t]
\begin{center}
\includegraphics[scale=0.6]{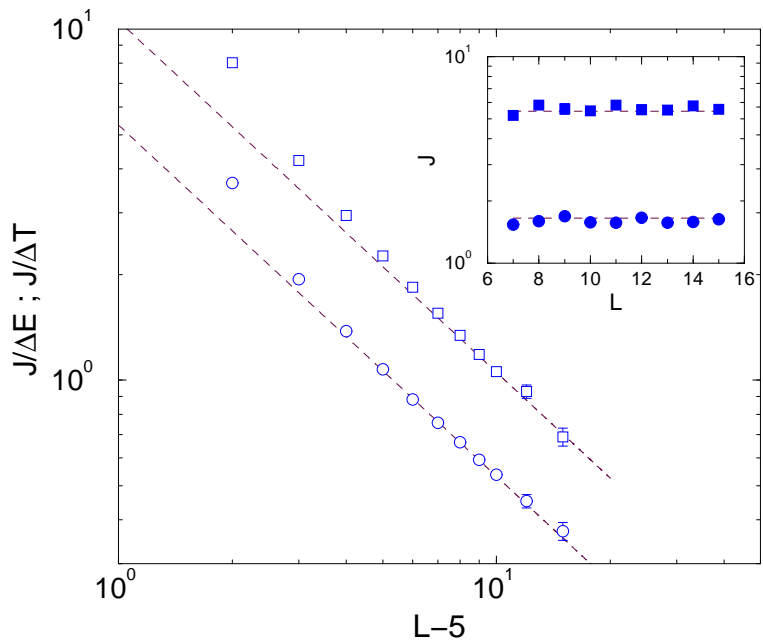}
\caption{\label{fig:4}
  Size dependence of the energy current in the chaotic chain with $T_{\rm l} =
  5$ and $T_{\rm r} = 50$.   We show $J/\Delta E$ (open circles) and $J/\Delta
  T$ (open  squares).  The dashed lines  corresponds to $1/\Delta  L$ scaling. 
  In the  inset, the size  dependence of the  energy current is shown  for the
  integrable (solid circles) and the intermediate (solid squares) cases.
 }
\end{center}
\end{figure}
 
In  fig.~\ref{fig:3} we  show the  energy profile  $\ave{H_n}$ for  an  out of
equilibrium  simulation   of  the  chaotic  chain.    In  all  non-equilibrium
simulations,  the temperatures  of the  baths were  set to  $T_{\rm  l}=5$ and
$T_{\rm r}=50$.  After an appropriate scaling the profiles for different sizes $L$
collapse to  the same curve.  More interesting, in  the bulk of the  chain the
energy profile is in very good approximation linear.
In  contrast, we  show  that  in the  case  of the  integrable  (inset I)  and
intermediate (inset II) chains, no energy gradient is created which is a characteristic
of ballistic transport.

We now define the local  current operators through the equation of continuity:
$\partial_t{H_n} =  \ii [ H,H_n] =  - (J_{n+1} - J_n)$,  requiring that
$J_n = -\ii [H_n,H_{n-1}]$.  Using eqs.  (\ref{eq:H_local}) and (\ref{eq:hamaut}) the
local heat current operators are explicitly given by
$ J_{n} = h_\z J\left(\sigma_{n-1}^\x-\sigma_{n+1}^\x\right)\sigma^\y_{n} $. 
In fig.~\ref{fig:4} we plot $J/\Delta E$ as  a function of the size $L$ of the
system  for sizes up  to $L=20$.   The mean  current $J$  is calculated  as an
average of $\ave{J_n}$ over time and  space $n$.
 The energy
difference was obtained from the energy profile as $\Delta E = \ave{H_{L-3}} -
\ave{H_2}$.  Two spins near each bath  have been discarded in order to be in
the bulk  regime. Since $\Delta L=L-5$  is an effective size  of the truncated
system, the  observed $1/\Delta L$  dependence confirms that the  transport is
normal (diffusive).  Moreover,  also  the  quantity  $J/\Delta  T$,  where  $\Delta  T  =
-1/\ave{H_{L-3}} + 1/\ave{H_2}$, shows the  correct scaling with the size $L$. 
On the other hand, in integrable and intermediate chains we have observed that
the average heat  current does not depend on the  size $J\propto L^0$, clearly
indicating the ballistic transport.
 
\section{Quantum relaxation and complexity in a toy model: Kicked Ising Chain}

\label{sect:statmech}
 
Let us now come back to the kicked Ising model (\ref{eq:hamiltonian}) and try to consider
some very elementary but fundamental questions considering its dynamics and non-equilibrium statistical
mechanics. 
Observing data of  fig.~\ref{fig:corr} in section (\ref{sect:losch}) one can conclude that the model perhaps displays
an interesting order to chaos, or non-ergodicity to ergodicity transition when the integrability breaking parameter
is increased. Now we would like to inspect this transition more closely, and in particular understand the
rate of relaxation to equilibrium in the ergodic and mixing case. We should stress right from the start that we are unable to prove
any non-trivial statements about the model, but we can provide many suggestive numerical experiments which can be performed
in a rather efficient way. We have learned in section \ref{sect:MPS} that it may be more fruitful to consider time
evolution in the operator algebra spaces instead of in the spaces of pure states. Let us go now a bit deeper into this subject.

For some related results on high-temperature relaxation in isolated conservative many-body quantum systems see e.g. 
Refs.\cite{fine,fabricius,lasinio,howard,ProsenPRE99qft}.

\subsection{Time automorphism}

Time automorphism  on unital quasi-local $C^*$ algebra (see e.g. \cite{BR} for introduction into the subject) 
${\cal A}_\Z$, $\aT : {\cal A}_\Z \to {\cal A}_\Z$ of an infinite KI lattice, 
for one period of the kick, can be explicitly constructed by the following observations.

Formally, for any $A\in {\cal A}_\Z$, $\aT A := U^\dagger A U$, where $U$ is given
by either (\ref{eq:floquet},\ref{eq:protocol},\ref{eq:protocol2}).
Let ${\cal A}_{[m,n]}$, with $m\le n$, denote a finite, local $4^{n-m+1}$ dimensional
algebra on a sub-lattice $[m,n] \subset \Z$, which is spanned by 
operators $\s{s_m}{m}\s{s_{m+1}}{m+1}\cdots\s{s_{n}}{n}$.
It is straightforward to prove that dynamics is {\em strictly local}
\begin{equation}
\aT : {\cal A}_{[m,n]} \to {\cal A}_{[m-1,n+1]}.
\label{eq:locality}
\end{equation}
In other words, the homomorphism $\aT$ (\ref{eq:locality}) is a simple nontrivial example of a quantum
cellular automaton as defined by Schumacher and Werner \cite{schumacherwerner}.

${\cal A}_\Z$ can also be treated as a Hilbert space with respect to the following
inner product $(A|B) = \omega(A^\dagger B)$, where $\omega(A)$ is a tracial state,
$\omega(A) = 2^{-(n-m+1)}\tr A$ for $A\in {\cal A}_{[m,n]}$.
This Hilbert space can be in fact considered as a 1D lattice of 4-level quantum 
systems (qudits with $d=4$) with the orthonormal basis $\Ket{\ldots,s_{-1},s_0,s_1,\ldots} \equiv 
\cdots\s{s_{-1}}{-1}\s{s_0}{0}\s{s_1}{1}\cdots$ 
labeled by an infinite sequence of 4-digits $s_j\in \Z_4$.
Restricting for a moment to a dimer lattice ${\cal A}_{[j,j+1]}$ we can write the adjoint 
action of a 2-qubit gate $W$ (\ref{eq:Wgate}) in terms of a $16\times 16$ unitary matrix
\begin{equation}
\aW_{j,j+1} \Ket{s_j,s_{j+1}} = W^\dagger \s{s_j}{j}\s{s_{j+1}}{j+1} W = 
\!\!\sum_{r_j,r_{j+1}\in\Z_4}\!\!\!\Ket{r_j,r_{j+1}}\aW_{(r_j,r_{j+1}),(s_j,s_{j+1})} 
\end{equation} 
where, very explicitly 
\begin{equation}
\aW_{(r_1,r_2),(s_1,s_2)} =  
\frac{1}{4}\tr \left[(\sigma^{r_1}\otimes\sigma^{r_2})  W^\dagger (\sigma^{s_1}\otimes\sigma^{s_2}) W\right].
\quad\quad
\end{equation}
We should note that the map is unital $\aW \Ket{0,0} = \Ket{0,0}$, and that due to anti-unitary symmetry of 
 KI model, the matrix $\aW$ is {\em real}. We extend the map $\aW_{j,j+1}$ to entire algebra ${\cal A}_\Z$ by
 $\aW_{j,j+1} (A\otimes B) = \aW_{j,j+1}(A)\otimes B$, for any $A\in {\cal A}_{[j,j+1]}$, 
 $B\in {\cal A}_{\Z - [j,j+1]}$.
 
Now, following the protocol (\ref{eq:protocol2}) we finish the construction of  the time automorphism as a 
string of right-to-left ordered 2-qudit (with $d=4$) gates
\begin{equation}
 \aT = \prod_{j\in\Z}^- \aW_{j,j+1}.
 \end{equation}
 Explicitly, for any local observable
 $A = \sum_{s_m,s_{m+1},\ldots,s_n} a_{(s_m,s_{m+1},\ldots, s_n)}\Ket{s_m,s_{m+1},\ldots,s_n} \in 
 {\cal A}_{[m,n]}$, we have the following 

\noindent {\bf Algorithm 1}:
\begin{enumerate}
\item Set an initial vector: $a^{(0)}_{(s_{m-1},s_m,\ldots,s_{n},s_{n+1})} = 
\delta_{s_{m-1},0} a_{(s_m,\ldots, s_n)} \delta_{s_{n+1},0}$.
\item For $k = 0,1,\ldots, n-m+1$:
\begin{equation}
\hspace{-1in} a^{(k+1)}_{(s_{m-1},s_m,\ldots,s_n,s_{n+1})} =
\sum_{r,r'\in\Z_4} \aW_{(s_{m-1+k},s_{m+k}),(r,r')}
a^{(k)}_{(s_{m-1},\ldots,s_{m+k-2},r,r',s_{m+k+1},\ldots,s_{n+1})}
\end{equation}
\item The result is
$\aT A = \sum_{s_m,s_{m+1},\ldots,s_n} 
a^{(n-m+2)}_{(s_{m-1},s_m,\ldots, s_{n+1})}\Ket{s_{m-1},s_{m},\ldots s_{n+1}} .$ 
\end{enumerate}
The algorithm produces exact result in $(n-m+2)4^{n-m+4}$ multiplications and about the same 
number of additions.
Let $\aP_{[m,n]} : {\cal A}_{\Z} \to {\cal A}_{[m,n]}$ denote a {\em linear orthogonal projector}, 
satisfying $\aP_{[m,n]}(A\otimes E) = (\one|E)A$ if $A\in{\cal A}_{[m,n]}$,$E\in{\cal A}_{\Z-[m,n]}$.
Let us define a truncated time evolution operator
$\aT_{[m,n]} = \aP_{[m,n]} \aT : {\cal A}_{[m,n]} \to {\cal A}_{[m,n]}$
which we can actually implement on a computer with a finite memory register.

The ``infinite-temperature'' time correlation function of (traceless) local quantum observables 
$A,B\in {\cal A}_{[m,n]}$ can be written as  $C_{BA}(t)=\Braket{B}{\aT^t A}$, where $t\in\Z$.
An interesting question is, up to what time $t$ the $C_{BA}(t)$ can be computed numerically exactly
with a finite computer register $[m-l,n+l]$ of $r=n-m+1+2l$ qudits of size $4^r$?
Due to locality (\ref{eq:locality}) of time homomorphism one can easily prove that
\begin{equation}
\Braket{B}{\aT^t A} = \Braket{B}{\aT_{[m-l,n+l]}^t A}, \quad {\rm for} \quad t \le 2l
\label{eq:excorloc}
\end{equation}
hence the correlation functions are computable exactly up to time $2l$, and as we shall
see later, truncated correlation function $C^{(r)}_{BA}(t)=\Braket{B}{\aT^t_{[m-r,n+r]} A}$
often well approximates $C_{BA}(t)$ even at later times, or even its asymptotic decay.

Let us continue our discussion by considering time evolution for {\em translationally
invariant extensive} (TIE) observables. Given some quasi-local observable $A\in{\cal A}_\Z$
we shall construct the corresponding TIE observable by a formal mapping,
$A \to \aF(A)=\sum_{x\in\Z}  \aS_x (A)$ in terms of lattice translation automorphisms 
$\aS_x :{\cal A}_\Z\to{\cal A}_\Z$,  $\aS_x (\s{s}{j}) = \s{s}{j+x}$.
The image of the entire quasi-local algebra under this mapping ${\cal Z} = \aF({\cal A}_{\Z})$,
is not a $C^*$ algebra,  but it is a linear space which can be again turned into a Hilbert space
with the following inner product
\begin{equation}
\BBraket{X}{Y} = 
\lim_{n\to\infty} \frac{1}{2n+1}\Braket{\aP_{[-n,n]}(X)}{\aP_{[-n,n]}(Y)}
\end{equation}
where the domain of projector $\aP_{[-n,n]}$ is extended to ${\cal Z}$ by continuity.
Orthonormal basis of ${\cal Z}$ is given by TIE observables
$Z_{(c_0,c_1,\ldots,c_{r-1})} = \aF( \s{c_0}{0}\s{c_1}{1}\cdots\s{c_{r-1}}{r-1})$,
for {\em orders} $r=1,2,\ldots$, and for uniqueness of notation, 
requiring $c_0,c_{r-1}\neq 0$. We shall interchangeably represent finite sequences
of 4-digits with integers, $(c_0,c_1,\ldots c_{r-1})\equiv
c=\sum_{j=0}^{r-1} c_j 4^j$. 
Let ${\cal Z}_r$ be $3\times 4^{r-1}$ dimensional
subspace spanned by TIE observables $Z_c$ with order $\le r$, i.e. for $c$ having at most 
$r$ base-4 digits, so we have an inclusion sequence 
${\cal Z}_1 \subset {\cal Z}_2 \ldots \subset {\cal Z}$.

Since time and space automorphisms commute $\aT \aS_x = \aS_x \aT$, 
one can immediately extend the time map onto the space of 
TIE observables, $\bT : {\cal Z} \to {\cal Z}$ by continuity.
Formally, we have $\bT \aF = \aF \aT$.
Furthermore, locality (\ref{eq:locality}) implies
\begin{equation}
\bT : {\cal Z}_r \to {\cal Z}_{r+2},
\end{equation}
so we can write a simple adaptation of Algorithm 1 for
explicit construction of a time map of an arbitrary finite-order TIE
observable  $Y = \sum_{0 < c < 4^r}^{c\neq 0\!\!\!\pmod{4}} y_c Z_c \in {\cal Z}_r$:
 
\noindent {\bf Algorithm 2}:
\begin{enumerate}
\item Take the following pre-image of the TIE observable
$ A = \sum_{0 < c < 4^r}^{c\neq 0\!\!\!\pmod{4}} y_c \Ket{4c} \in 
{\cal A}_{[1,r]}$, namely $\aF(A) = Y$.
\item Compute $a_c$ of
$\aT(A) = \sum_{0 \le c < 4^{r+2}} a_c \Ket{c} \in {\cal A}_{[0,r+1]}$ 
according to {\bf Algorithm 1}.
\item
Transforming back to ${\cal Z}$ the result reads
\begin{equation}
\hspace{-2cm}\bT(Y) = \aF (\aT(A)) = \!\!\!\sum_{0 < c < 4^{r+2}}^{c\neq 0\!\!\!\pmod{4}}\!\!\!
y'_c Z_c,\;\;
y'_c = \left\{\begin{array}{ll}
a_c + a_{4c} + a_{16c}, & c < 4^{r};\\
a_c + a_{4c}, & 4^{r} \le c < 4^{r+1};\\
a_c, & c \ge 4^{r+1}.
\end{array}\right.
\end{equation} 
\end{enumerate}
Let us further define the {\em natural truncations} of TIE space to order $r$, 
$\bP_r : {\cal Z} \to {\cal Z}$
as orthogonal projections 
$\bP_r(X) = \sum_{0 < c < 4^r}^{c\neq 0\!\!\!\pmod{4}} Z_c \BBraket{Z_c}{X}$,
and truncated time evolution operators $\bT_r = \bP_r \bT : {\cal Z}_r \to {\cal Z}_r$,
which are naturally implemented on a computer by simply truncating overflowing coefficients $y'_c$.

Physically interesting question now concerns computation of time correlation functions
between a pair of finite order (say $q$) TIE observables $X,Y\in {\cal Z}_q$, namely
$C_{YX}(t) = \BBraket{Y}{\bT^t X}$, for example in fig.~\ref{fig:corr} we have shown the case
of $X=Y=Z_3$. As a consequence of locality (\ref{eq:locality}), and translational invariance,
we find that the truncated evolution on ${\cal Z}_r$, reproduces correlation functions exactly
\begin{equation}
\BBraket{B}{\bT^t A} = \BBraket{B}{\bT_r^t A},\quad{\rm for}\quad t \le r - q.
\end{equation}
Few remarks are in order: 
(i) Truncated translationally invariant time evolution $\bT_r$ is perhaps more natural object to study than truncated local
time evolution $\aT_{[m,n]}$, for a simple reason that the truncation $\bP_r$ commutes with a
shift $\aS_x$, while $\aP_{[m,n]}$ does not.
(ii) A space ${\cal Z}$ can be identified with translationally invariant linear functionals over ${\cal A}_\Z$, 
namely $X(A) = \BBraket{X}{\aF A}$, $X\in{\cal Z}$, $A\in{\cal A}_\Z$. 
We have $(\bT^\dagger X)(A) = X(\aT A)$ and $(\bT X)(A) = X(\aT^\dagger A)$
where Hermitian adjoint maps $\bT^\dagger$ and $\aT^\dagger$ simply correspond to {\em time reversed} dynamics.
(iii) Convex subspace of positive translationally invariant functionals ${\cal W}\subset{\cal Z}$ is an interesting
invariant subspace of physical states, $\bT {\cal W} \subseteq {\cal W}$.

\subsection{Relaxation and quantum Ruelle resonances}

In classical mechanics of chaotic systems one typically observes that states (phase-space
densities) develop small details in the course of time evolution at an exponential average rate.
Consequently, introducing a small stochastic noise of strength $\epsilon$ to 
Perron-Frobenius operator (PFO, i.e. Liouvillian propagator for discrete time dynamics),
makes it non-unitary and shifts its spectrum inside the unit circle. Typically, the effect of noise is equivalent to an ultraviolet cutoff - truncation of PFO - at the Fourier scale $k \sim 1/\epsilon$,
and often the leading eigenvalues of truncated PFO  - the so-called Ruelle resonances - remain frozen inside the unit circle in
the limit $\epsilon\to 0$ (or $k\to \infty$) \cite{ktopcl}.
For a general introduction to relaxation phenomena in classical Hamiltonian dynamics see e.g. \cite{gaspard}.

Let us now draw some some analogies with our quantum setting.
We have seen that that the evolution $\bT$ somehow most closely resembles Liouvillian
evolution of classical Hamiltonian dynamics. In Hilbert space topology, operator $\bT$ is unitary 
and its spectrum lies on a unit circle, just like in the case of classical PFO.
However, truncated $(3\times 4^{r-1}) \times (3\times 4^{r-1})$ matrices $\bT_r$ represent natural ``ultraviolet''  cutoff truncations
for increasing orders $r$. Let us check numerically if some eigenvalues of these matrices remain frozen when 
$r\to \infty$.

\begin{figure}
\centerline{\includegraphics[width=11cm,angle=-90]{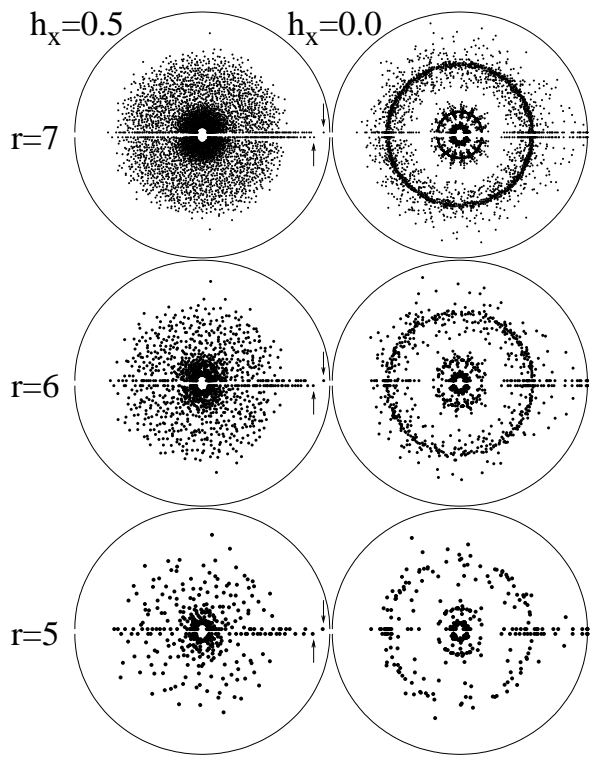}}
\caption{The spectra of truncated transfer operators $\bT_r$, for $r=5,6,7$ in strongly 
non-integrable case $J=0.7,h_\x=0.5, ,h_\z=1.1$ (left) and integrable case $J=0.7, h_\x=0.0,h_\z=1.1$ (right), 
lying inside complex unit 
circle (thin arcs). The points in upper/lower unit semi-disks correspond to positive/negative
parity $\bR Z_{(c_0,c_1,\ldots,c_{l-1})} = Z_{(c_{l-1},\ldots,c_1,c_0)}$
eigenvectors. Arrows point at converged positions of the leading eigenvalue $e^{-q_1}$.}
\label{fig:s}
\end{figure}

Indeed, as we demonstrate in fig.~\ref{fig:s}, we find several eigenvalues which converge as $r$ increases in the case of strongly non-integrable
(quantum chaotic) case, with a gap between an eigenvalue of maximal modulus and the unit circle,
 whereas in the integrable case a set of $r$ eigenvalues touch the unit circle (actually eigenvalue 1 is
 $r$-fold degenerate).
Numerical results suggest the following speculative conclusions.
Let $e^{-q_n}$ be the converged (frozen) eigenvalues of $\bT_r$, and
$\{\Theta^{\rm R}_n\}$, $\{\Theta^{\rm L}_n\}$
the corresponding {\em right} and {\em left} eigenvectors, respectively.
Then for arbitrary pair $X,Y\in{\cal Z}$, the time correlation function can be expressed in terms
of spectral decomposition (see e.g. \cite{gaspard})
\begin{equation}
C_{YX}(t) \sim \sum_n w_n e^{-q_n t},
\quad w_n = 
\frac{\BBraket{Y}{\Theta^{\rm R}_n}\BBraket{\Theta^{\rm L}_n}{X}}{\BBraket{\Theta^{\rm L}_n}{\Theta^{\rm R}_n}}.
\label{eq:corr}
\end{equation}
The above relation is the contribution of the point spectrum and is exact if the spectrum
is pure-point. However, in classical cases one may quite typically have various singular components and
branch cuts \cite{gaspard}.
Note that the denominator 
$\BBraket{\Theta^{\rm L}_n}{\Theta^{\rm R}_n}$ is finite although both vectors should have infinite $l^2$ norm 
$\BBraket{\Theta^{\rm L}_n}{\Theta^{\rm L}_n}=\infty$, $\BBraket{\Theta^{\rm R}_n}{\Theta^{\rm R}_n}=\infty$,
for any eigenvalue away from the unit circle, $\re q_n\neq 0$.

There is a simple relation between the spectrum of PFO and the ergodic properties of
dynamics:
(i) If there is a {\em spectral gap},
i.e. there exists $\lambda > 0$ such that for all $n$, $|e^{-q_n}| \le \exp(-\lambda) < 1$, then dynamics is {\em exponentially mixing}, 
$C_{YX}(t) \le C \exp(-\lambda t).$
(ii) If some eigenvalues are on the unit circle, meaning that the corresponding eigenvector
coefficients should be in $l^2$, then the system is {\em non-mixing}
since there are correlation functions which do not decay.
(iii) If some eigenvalues are at $1$ then the system is {\em non-ergodic} since the
correlation functions may have non-vanishing time-averages. If $Q_n$ is a complete
set of orthonormalized eigenvectors corresponding to eigenvalue $1$, 
$\BBraket{Q_n}{Q_m}=\delta_{n,m}$ (and note that since we are on the
unit circle: $Q^{\rm R}_n = Q^{\rm L}_n$) then
\begin{equation}
D_X := \overline{C_{XX}(t)} = \sum_n | \BBraket{X}{Q_n}|^2.
\end{equation}
The latter (iii) happens in generic completely integrable quantum lattices, where $Q_n$
correspond to an infinite sequence of conservation laws \cite{integrable}. Furthermore, 
we have a strong numerical evidence that also in certain non-integrable quantum 
lattices \cite{ProsenJPA98}, and also in KI model \cite{ProsenPRE02}, one has a regime where 
few normalizable ('pseudo-local') but not local (like in integrable models) 
conservation laws exist. This situation we call the regime of {\em intermediate} dynamics and is 
characterized by a non-vanishing {\em stiffness} $D_X \neq 0$ signalling {\em ballistic transport}.

\begin{figure}
\centerline{\includegraphics[width=3in,angle=-90]{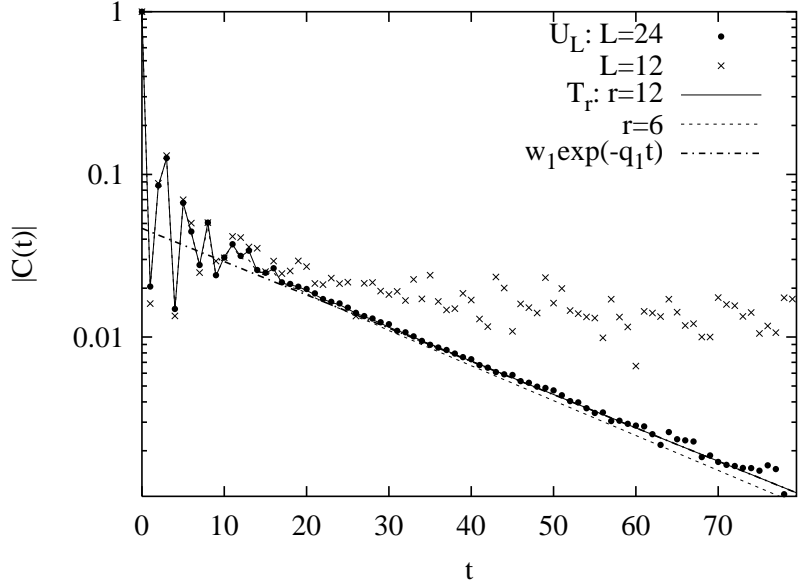}}
\caption{Correlation function of the transverse magnetization $C(t) =\BBraket{M}{M(t)}$, in the mixing 
case $J=0.7,h_\x=0.5,h_\z=1.1$, computed from finite system dynamics for different 
sizes $L$ (symbols), and from truncated adjoint propagators $\bT_r$ of infinite systems (curves)
for different truncation orders $r$. The chain line indicates the asymptotics based on the leading
quantum Ruelle resonance.}
\label{fig:c}
\end{figure}

In figure \ref{fig:c} we compare the time autocorrelation function of the 
transverse magnetization
$ M = \sum_{j\in\Z} \s{\z}{j} = Z_{3},$
computed in three different ways: (1) from exact time evolution 
$C_L(t) = \frac{1}{L}\ave{M U^{-t}_{L} M U^t_{L}}$
on a finite lattice of length $L$ with periodic boundary conditions,
(2) iteration of truncated TA matrix on infinite lattice 
$C_r(t)=\BBraket{M}{\bT_r^t M},$
and (3) asymptotics based on (few) leading eigenvalue resonance(s) 
(using formula (\ref{eq:corr}) in terms of $q_n$ and $w_n$.).

\begin{figure}
\centerline{\includegraphics[width=8cm,angle=-90]{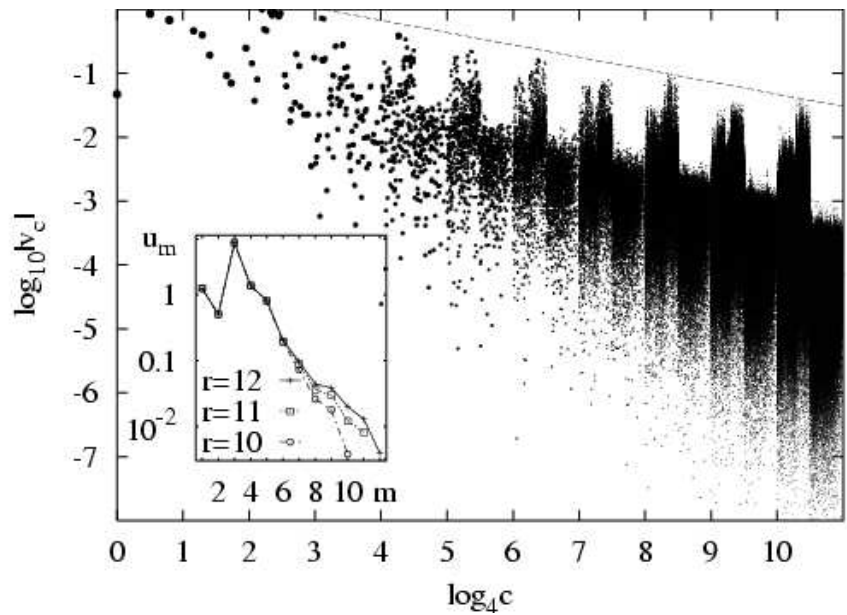}}
 \caption{
Eigenvector $\Theta^{\rm R}_1=\sum_c v_c Z_c$, 
of the leading eigenvalue (closest to unit circle) has statistically 
selfsimilar structure, when expanded in $Z_c$.
We plot the modulus of coefficients $v_c$ (in log scale) of the right eigenvector versus
the log (in base 4) of the integer code $c$. Dashed line indicates power
law scaling $c^{-\nu}$ with slope $\nu = 0.32$. In the inset we plot partial scalar products
$u_m = \sum_{4^{m-1} \le c < 4^m-1}^{c\neq 0\!\!\pmod{4}} 
\BBraket{\Theta^{\rm L}_1}{Z_c}\BBraket{Z_c}{\Theta^{\rm R}_1}$ 
with the corresponding left eigenvector within fixed orders $m$.
 }
\label{fig:v}
\end{figure}

\begin{figure}
       \centering
       \includegraphics[width=8cm,angle=0]{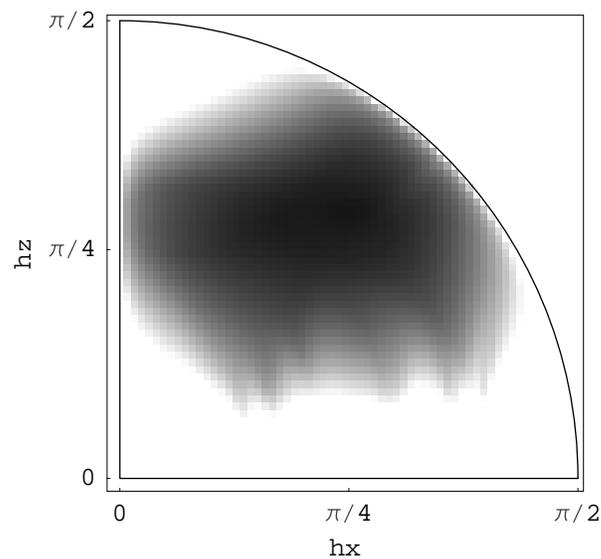}
       \caption{Two dimensional numerical phase diagram for kicked Ising lattice at cutoff order $r=7$. Gray level indicates the
       spectral gap $\log \Delta$ of $\bT_r$ as a function of $h_\x$ and $h_\z$ at fixed $J=0.7$.  White regions correspond to 
       $\Delta < 10^{-6}$.}
       \label{fig:phasediagram}
\end{figure}

We note that the leading eigenvalue and eigenvector of truncated TA $\bT_r$
can most efficiently be computed using our Algorithm 2 as a key step of 
an iterative {\em power-method}. In this way we were able to perform calculations 
of the leading Ruelle resonances up to $r=15$ in contrast to full diagonalization of truncated matrices $\bT_r$ 
which were feasible only up to $r=7$. Let us observe the structure of the eigenvector coefficients 
$v^{\rm L,R}_c = \BBraket{Z_c}{\Theta_n^{\rm L,R}}$ corresponding to the leading eigenvalue. Numerical
results (see fig.~\ref{fig:v}, see also subsect.\ref{sect:scaling} later) strongly suggest self-similar behaviour upon multiplying the
code $c$ by $4$ which is a consequence of the fractal structure of the transfer matrix 
(see illustration in Ref.\cite{ProsenRu,Prosen04}).  
\begin{figure}
 
\hfil\includegraphics[width=8.0cm,angle=-90]{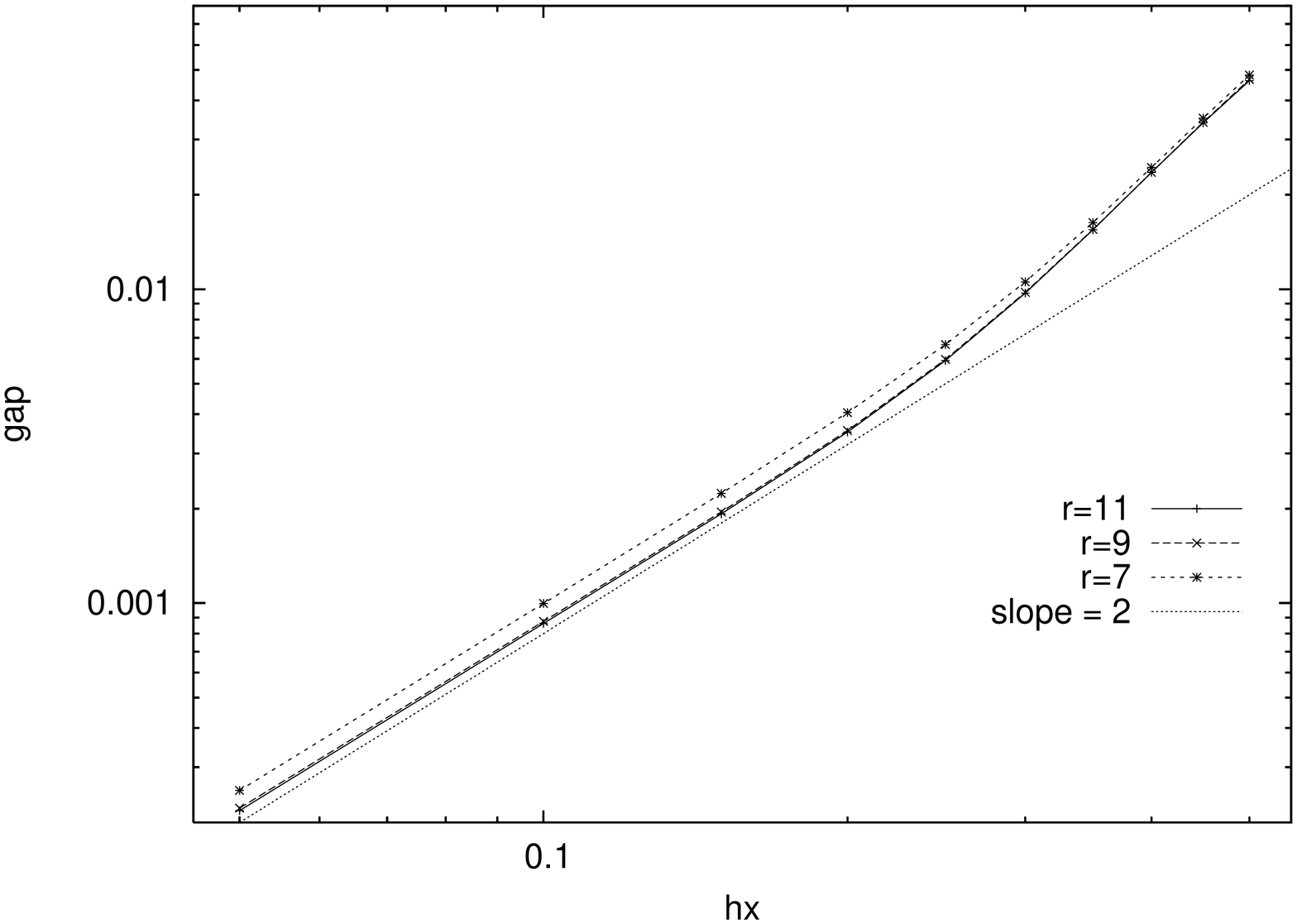}

\vspace{-7.5cm}\hspace{4cm}
\includegraphics[width=4.0cm,angle=0]{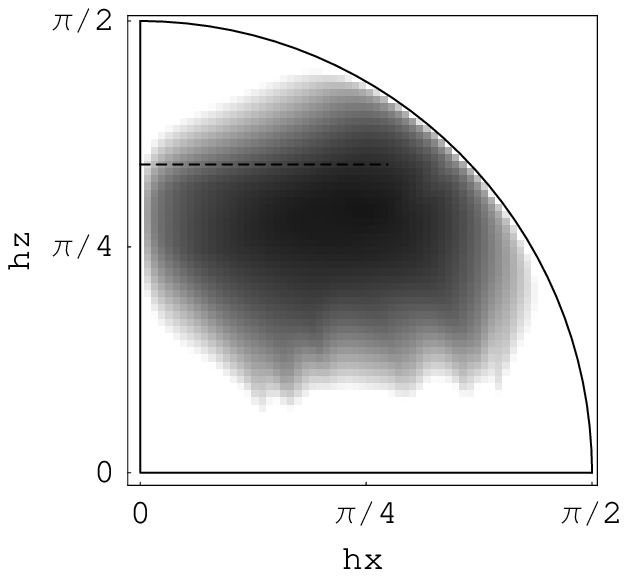}  

\vspace{3.5cm}

\caption{Type I transition: Spectral gap of 
$\bT_r$  as a function of $h_\x$ at $J=0.7,h_\z=1.1$ at different truncation orders $r$. In the inset we indicate a line
of transition in 2D phase diagram.
}
\label{fig:typeI}
\end{figure}
\begin{figure}
 
\hfil\includegraphics[width=8.0cm,angle=-90]{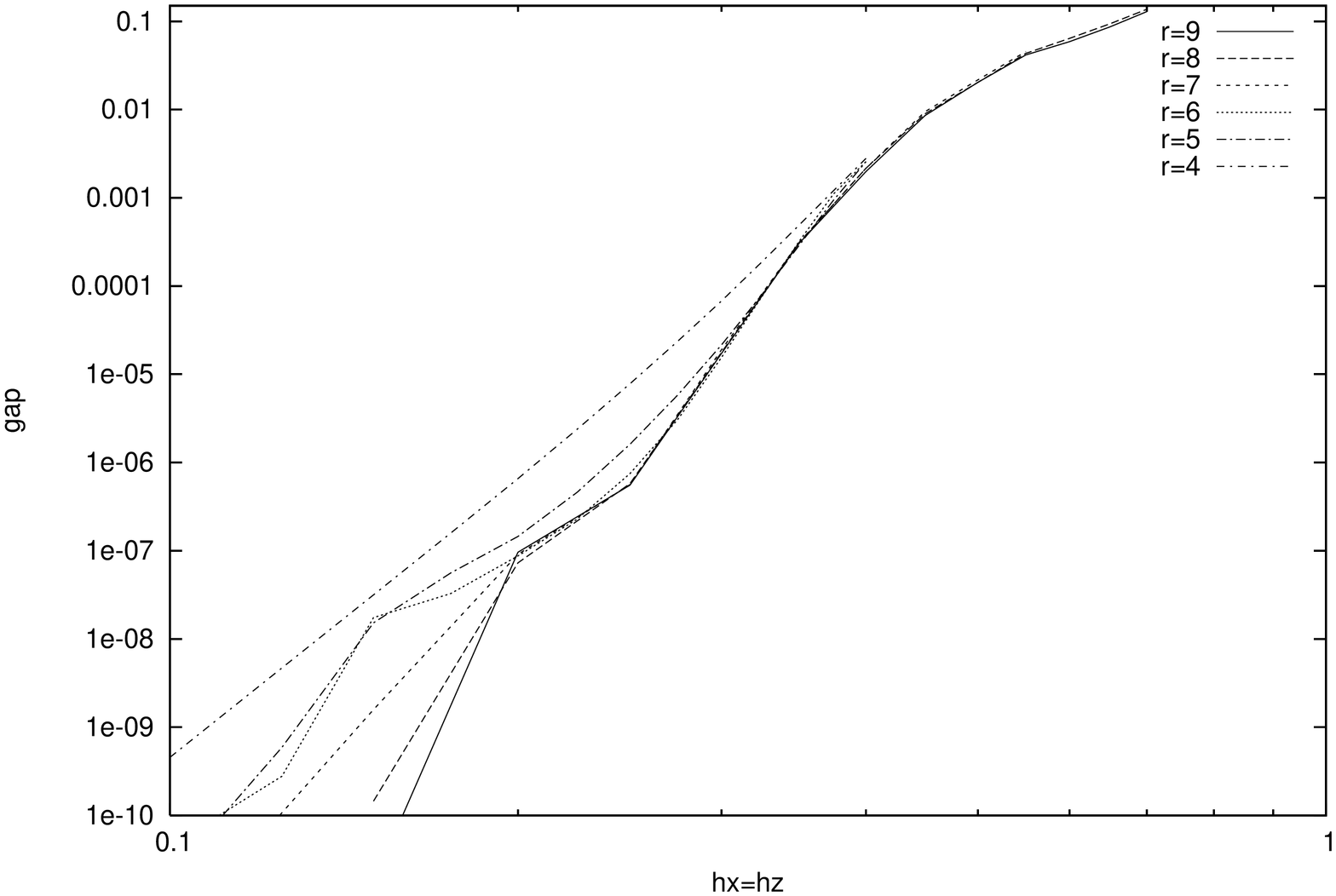}

\vspace{-4.5cm}\hspace{8.5cm}
\includegraphics[width=4.0cm,angle=0]{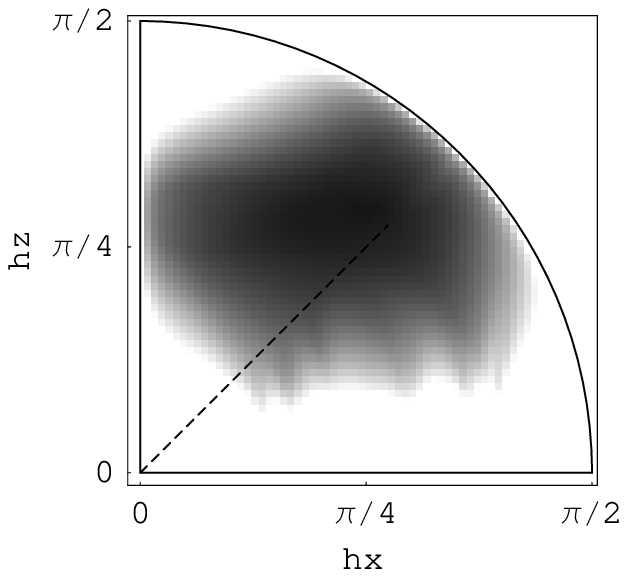}  

\vspace{0.5cm}

\caption{Type II transition:
Type I transition: Spectral gap of 
$\bT_r$  as a function of $h_\x=h_\z$ at $J=0.7 $ at different truncation orders $r$. In the inset we indicate a line
of transition in 2D phase diagram.
}
\label{fig:typeII}
\end{figure}

The stiffness $D_X$ and  the spectral gap $\Delta=|1-e^{-q_1}|$ may be considered as {\em order parameters}, characterizing
a particular kind of dynamical phase transition, namely the transition from non-ergodic dynamics -  ordered phase, 
where $D_X \neq 0$ for a typical $X$\footnote{Any observable $X$ which is not orthogonal to all
conservation laws $Q_n$, i.e. normalizable eigenvectors of $\bT$ with eigenvalue $1$.}, 
or $\Delta=0$, to an ergodic and mixing dynamics - disordered phase,
where $D_X = 0$, for all traceless $X\in{\cal Z}$, or $\Delta > 0$.
We know that KI model is non-ergodic in integrable regimes. Let us consider a fixed transverse field case $h_\x=0$, and start to switch on a small amount of longitudinal field $h_\x$.
The interesting question is whether the transition happens for infinitesimal integrability breaking parameter $h_\x$ in TL,
or at a finite - {\em critical} field. In fig.~\ref{fig:phasediagram}, we fix $J=0.7$ and plot a two dimensional phase diagram of the 
spectral gap $\Delta(h_\x,h_\z)$. It is clear that we have different behaviours in different regions of parameter space, 
for example we identify two:
(i) Type I transition: if the transverse field is roughly on the interval $h_\z \in [0.7,1.2]$, then the spectral gap opens in the fastest possible
manner which is allowed by a $h_\x \to -h_\x$ symmetry and the analyticity of the problem, namely $\Delta \propto h_\x^2$. See fig.~\ref{fig:typeI}.
(ii) Type II transition: if the initial transverse field $h_\z < 0.7$, or $h_\z > 1.2$, then the gap opens up in a much 
more abrupt - perhaps a discontinuous way. We give an example by scanning the diagonal transition, i.e. putting $h_\x=h_\z$ and
increasing $h_\x$ from zero. Numerical results, shown in fig.\ref{fig:typeII} cannot be made fully conclusive, but they are
not inconsistent with a conclusion that an abrupt transition to ergodic behaviour takes place at $h_\x = h_\z \approx 0.3$.

\subsection{Translationally invariant conservation laws as matrix product operators}

There is another possibly interesting way of characterizing the transition, i.e. in terms
of pseudo-local translationally invariant conservation laws \cite{ProsenJPA98}.
Such conservation laws are the square normalizable elements $Q\in {\cal Z}$, which are mapped
onto themselves under the dynamics $\bT(Q) = Q$. We shall first make a non-trivial variational MPO ansatz for 
elements of ${\cal Z}$, namely let us take an auxiliary vector space $\C^D = \C^{D_1}\oplus\C^{D_2}\oplus\C^{D_3}$, where
$D=D_1+D_2+D_3$. 
Then any operator $Q$, which is formally written in terms of MPO on the {\em infinite} spin chain 
\begin{equation}
Q = \sum_{\ldots s_{-1} s_0 s_1 \ldots\in\Z_4}
(\vec{a}_{\rm L}\cdots A^{s_{-1}} A^{s_0} A^{s_1} \cdots\vec{a}_{\rm R})\cdots 
\sigma_{-1}^{s_{-1}} \sigma_0^{s_0} \sigma_1^{s_1} \cdots 
\label{eq:TIMPO}
\end{equation}
where $A^s \in \C^{D\times D},\vec{a}_{\rm L},\vec{a}_{\rm R}\in\C^D$ have the block matrix form (for $k=1,2,3$):
\begin{equation}
A^0 \!=\! 
%\begin{pmatrix}
\pmatrix{
1 & 0 & 0 \cr
0 & 1 & 0 \cr
0 & 0 & E_0 \cr
%\end{pmatrix}
}\!,
\;
A^{k} \!=\!
%\begin{pmatrix}
\pmatrix{
0 & * & * \cr
0 & 0 & 0 \cr
0 & * & E_k \cr
%\end{pmatrix}, 
}\!,
\; 
\vec{a}_{\rm L}\!=\!\pmatrix{* \cr 0 \cr 0\cr},
\;
\vec{a}_{\rm R}\!=\!\pmatrix{0 \cr * \cr 0 \cr}\!,
\label{eq:MPOdata}
\end{equation}
represents a translationally invariant pseudo-local operator, i.e. an element of ${\cal Z}$,
provided that $|| E_0 || < 1$ and $||E_1||^2 + ||E_2||^2 + ||E_3||^3 < 1$, where $||.||$ is a spectral
matrix norm and $*$'s stand for arbitrary matrices/vectors. 
Of course, converse cannot be generally true, not any element of ${\cal Z}$ can be written as 
MPO (\ref{eq:TIMPO}) with finite $D$, but still there are elements of the form (\ref{eq:TIMPO},
\ref{eq:MPOdata}) 
which are not in ${\cal Z}_r$ for {\em any} finite $r$. 

\begin{figure}
       \centering
       \includegraphics[width=8cm,angle=-90]{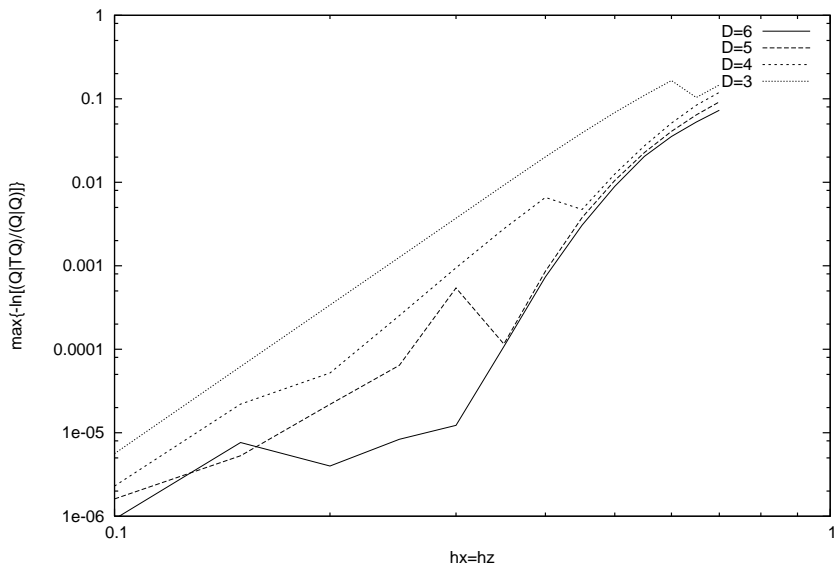}
       \caption{Optimized fidelity (\ref{eq:fidQ}) - in log scale - for approximate conservation laws within MPO spaces of different fixed  dimensions $D$ (indicated in the
       legend) for an infinite KI lattice along the diagonal type II transition with $h_\x=h_\z$ and $J=0.7$. 
       A   simple stochastic search has been
       used to maximize fidelity $F(Q)$. 
       Note the similarity with the gap curve shown in fig.\ref{fig:typeII}.}
       \label{fig:conslaw}
\end{figure}

There exist a straightforward algorithm which performs time evolution on
MPO data (\ref{eq:MPOdata}), namely $\bT (Q)$ is also of the
form (\ref{eq:TIMPO},\ref{eq:MPOdata}) with dimension $D' \le 2 D$.
We shall now make the following simple numerical experiment. Let us fix $D$, setting $D_1=D_2=1$
representing the simplest $\bT$-invariant subclass of (\ref{eq:TIMPO},\ref{eq:MPOdata}),
and optimize (maximize) the fidelity-like quantity
\begin{equation}
 F(Q) = \frac{|\BBraket{Q}{\bT Q}|}{\BBraket{Q}{Q}},
 \label{eq:fidQ}
\end{equation}
within this class of operators. Let us write an operator which maximizes $F(Q)$ for a given $D$ as $Q_D$. Note that $1-F(Q_D)$ gives a strong-topology measure of conservation of observable
$Q_D$ in one step of time evolution, so $F(Q_D) = 1$ only for exact conservation laws. 
Increasing $D$ may improve fidelity, if pseudo local conservation
laws exist to which $Q_D$ may converge, however in ergodic and mixing situation where no exact pseudo-local
conservation laws exist, increasing $D$ should have no significant effect to fidelity $F(Q_D)$.
This is exactly what we observe in KI model following a line of type II transition (see fig.~\ref{fig:conslaw}).
 
\subsection{Operator-space entanglement measures and complexity of time evolution}

\label{sect:opent}

Numerical results of section \ref{sect:MPS} suggested that operator space entanglement measures can be used
to characterize the complexity of time evolution, namely the minimal required rank $D_\epsilon$ of MPO ansatz is simply related
to entanglement entropy of a time-evolving local observable, which is interpreted as a Hilbert space
vector. In order to make things as simple and precise as possible we go back to the time evolution 
automorphism $\aT$ over the
quasi-local spin algebra ${\cal A}_\Z$. Let us take some truncation order $r = 2n+1$ and consider the truncated map 
$\aT_{[-n,n]}$. Starting with local operators on a single site ${\cal A}_{[0,0]}$ this truncated map is exact up to time 
$t=n$.
Writing time evolved observable at any instant of time as $A(t) = \aT^t_{[-n,n]} A(0) = 
\sum_{\ve{s}} a^{(t)}_{(s_{-n},\ldots s_n)} \Ket{s_{-n},\ldots,s_n}$
in terms of a `wave-function' $a^{(t)}_{(s_{-n},\ldots,s_n)}$, and partitioning a sub-lattice at $m, -n < m \le n$, as
$[-n,n] =  [-n,m-1] \cup [m,n]$, we can define $4^{n-m+1} \times 4^{n-m+1}$ reduced super-density matrix as
\begin{equation}
R^{(m,n)}_{(s_{m},\ldots s_n),(s'_{m},\ldots,s'_n)}(t) = 
\!\!\!\sum_{s_{-n},\ldots,s_{m-1}}\!\!\! a^{(t)}_{(s_{-n},\dots,s_{m-1},s_{m},\ldots s_n)} a^{(t)*}_{(s_{-n},\dots,s_{m-1},s'_{m},\ldots s'_n)}.
\label{eq:Rmatrix}
\end{equation}
Since $A(t)$ is interpreted as a `{\em pure state}', namely a vector from ${\cal A}_{[-n,m-1]}\otimes{\cal A}_{[m,n]}$,
the operator space entanglement is most simply characterized either by Von Neuman entropy 
$S^{(m,n)}(t) = -\tr R^{(m,n)} \log R^{(m,n)}$ or linear entropy $S^{(m,n)}_2(t) = -\log P^{(m,n)}(t)$
where $P^{(m,n)}(t) = \tr [R^{(m,n)}(t) ]^2$ is a purity of reduced super-density matrix.

\begin{figure}
       \centering
       \includegraphics[width=8.0cm,angle=-90]{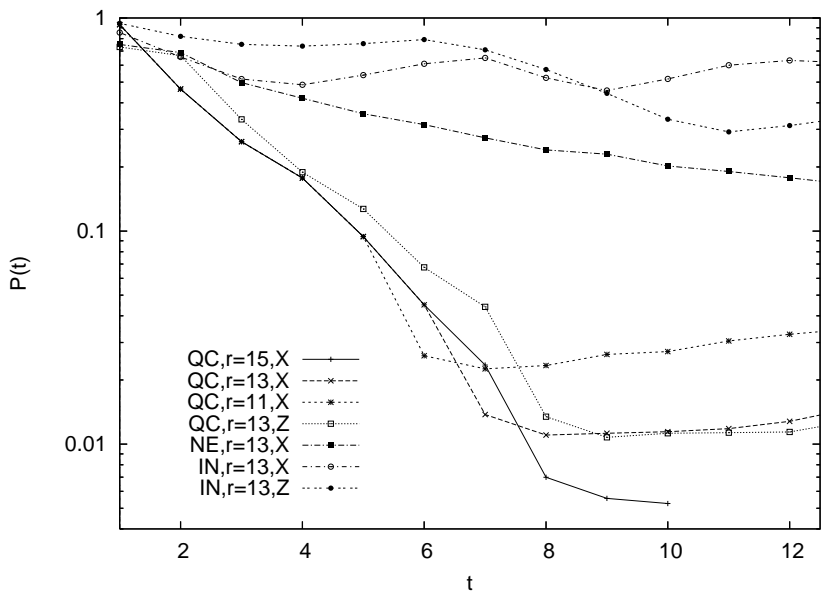}
       \caption{Purity $P^{(0,n)}(t)$ of reduced operator space density matrix for KI lattice, with different
       cutoff sizes $r=2n+1=15,13,11$, and for two initial centered local operators $X=\s{\x}{0}$
       and $Z=\s{\z}{0}$. Note exponential decay of purity for QC case and saturation or slow decay 
       of purity for non-ergodic cases (IN, NE).}
       \label{fig:pursuperrho}
\end{figure}

In fig.~\ref{fig:pursuperrho} we plot the entanglement purity $P^{(0,n)}(t)$ -- for close to symmetric bipartition 
$m=0$ where the entanglement is expected to be maximal -- for three different characteristic cases of KI model, which will
be in the following referred to as: {\em quantum chaotic} (QC), $J=0.7, h_\x = 0.9, h_\z = 0.9$,
{\em integrable} (IN), $J=0.7, h_\x=0, h_\z=0.9$, and {\em non-ergodic} (NE) non-integrable case,
$J=0.7,h_x=0.2,h_\z=0.2$. We find that in QC case purity decreases exponentially 
$P^{(0,n)}(t) = \exp(-h_{\rm q}t)$, meaning
$S_2(t) = h_{\rm q}t$, where the exponent $h_{\rm q}$ is independent of the initial observable $A(0)$ and asymptotically independent of $r$. On the other hand, in IN case $P^{(0,n)}(t)$ does not decay at all so the resulting dynamical entropy
$h_{\rm q}=0$, whereas in NE case $P^{(0,n)}(t)$ decays slowly, likely slower than exponentially.

\subsection{Scaling invariance and the problem on semi-infinite lattice}

\label{sect:scaling}

The eigenvectors of $\bT$ corresponding non-unimodular eigenvalues seem to exhibit a certain scaling invariance
(fig. ~\ref{fig:v}). Here we would like to explore this property in a little bit more detail. 
For that purpose we again explore the map $\aT_{[-n,n]}$ on the quasi-local algebra ${\cal A}_{[-n,n]}$  since the representation
of dynamics is conceptually simpler (Algorithm 1) than dynamics $\bT_{2n+1}$ on ${\cal Z}_{2n+1}$.
In particular, it is worth to mention that if one traces out an additional qudit after each time step, then the dynamics
is exact on a {\em closed} set of $4^{n-m+1}\times 4^{n-m+1}$ super-density matrices and has a simple explicit form in terms of a completely positive matrix map
\footnote{When labeling tensor product matrix elements we shall always follow a convention that left factors
are labelled with less significant digits, namely $(A\otimes B)_{j+ j'd,k+k'd} = A_{j,k} B_{j',k'}$.}
\begin{eqnarray}
&& R^{(m+1,n+1)}(t+1) = \tr_0\left\{T_{n-m}  \left[R^{(m,n)}(t) \otimes E_{00}\right] T^\dagger_{n-m}\right\}, 
\label{eq:rdynamics1} \\
&& T_{r} = \prod_{0\le k \le r}^- \one_4^{\otimes k}\otimes W\otimes \one_4^{\otimes (r-k)} ,
\label{eq:rdynamics2}
\end{eqnarray} 
namely no truncation is needed since at time $t+1$ we are describing an exact observable on ${\cal A}_{[-n-1,n+1]}$.
We write $(\tr_0 R)_{j,k} := \sum_{s=0}^3 R_{s+4j,s+4k}$ for tracing out the least significant qudit, and
$E_{00}=\Ket{0}\Bra{0}$ is an elementary $4\times 4$ projector.  Note that $T_{n-m}$ is just a matrix of 
$\aT_{[m,n]}$ in the canonical basis $\Ket{\ve{s}}$.
It is rather trivial to exactly solve this dynamics for a small finite $n-m$, however this does not yield physically very
useful information about the KI dynamics. One would wish to study the correct TL by first taking $n\to\infty$, and only
after that $m,t\to\infty$, however this task seems almost computationally intractable. Still, we were able to make some
modest numerical experiments exploring this question, suggesting that for sufficiently strong integrability
breaking (say QC case) the asymptotic matrix $R^{(m,\infty)}(\infty)$ has a remarkable scale invariance if we 
coarse-grain it by tracing over its $4\times 4$ blocks:
\begin{equation}
R^{(m+1,\infty)}(\infty) = \tr_0 R^{(m,\infty)}(\infty) = \zeta R^{(m,\infty)}(\infty),
\label{eq:scaling}
\end{equation}
where $\zeta$ is some scaling factor.
This seems to be true for both orders of the limits $t\to\infty$, $n\to\infty$, although better
numerical results have been obtained for the `incorrect' limit, namely letting the number of iterations $t\to\infty$
for a finite register size $n$, and then checking the convergence of results with increasing $n$.

\begin{figure}
       \centering
       \includegraphics[width=8.0cm,angle=-90]{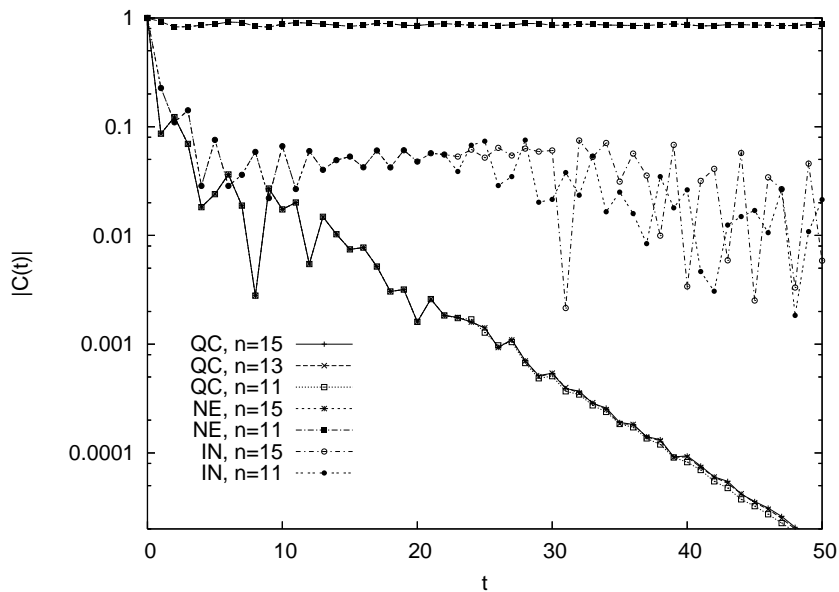}
       \caption{Decay of correlations $C^+(t) = \ave{\sigma_0^{\rm x}(t) \sigma_0^{\rm x}}$
       for semi-infinite KI lattice, with different cutoff sizes $n=15,13,11$, and for different
       cases (QC, NE, IN), all indicated in the figure. Note that the two curves for NE case
        are practically overlapping, and that all curves for different $r$'s are exactly overlapping until
        $t=2 r_{\rm smaller}$.}
       \label{fig:cfsc}
\end{figure}

Before discussing numerical results we note another useful observation. 
Let us define and briefly study KI chain on a {\em semi-infinite}
lattice $\Z_+=[0,\infty]$, with the Hamiltonian (\ref{eq:hamiltonian}) for $L=\infty$ and open boundary condition on the
left edge. Now we consider a quasi-local algebra ${\cal A}_{\Z_+}$. 
The time automorphism is again strictly local  
$\aT^+ : {\cal A}_{[0,n]} \to {\cal A}_{[0,n+1]}$, and can be written as a semi-infinite product
$\aT^+ = \prod_{j\in \Z_+}^- \aW_{j,j+1}$. In the definition of the truncated time map
$\aT^+_{[0,n]} = \aP_{[0,n]} \aT^+ : {\cal A}_{[0,n]} \to {\cal A}_{[0,n]}$ truncation is needed only on the right edge.
Note that due to this property, simulation of local observables (localized near the edge of the lattice)
is twice as efficient than on doubly-infinite lattice, meaning that with the same size of computer register
one can exactly simulate for twice longer times.
For example, computation of correlation functions $C^+_{BA}(t) := \Braket{B}{[\aT^+]^t A}$ is exact
\begin{equation}
\Braket{B}{[\aT^+]^t A} = \Braket{B}{[\aT^+_{[0,n]}]^t A},\quad {\rm for}\;\; t \le 2(n-q),
\quad {\rm if}\;\; A,B\in {\cal A}_{[0,q]}.
\end{equation}
Numerically inspecting the correlation functions of a simple local spin $A=B=\s{\x}{0}$ in fig.~\ref{fig:cfsc} we find
clear asymptotic exponential decay for QC case, whereas for IN and NE cases we find non-vanishing plateaus in the correlation function, 
i.e. non-vanishing stiffness $D^+ :=\overline{C^+} \neq 0$ which signals non-ergodicity and existence of local (for integrable
cases) and pseudo-local (for non-ergodic and non-integrable cases, e.g. NE) conservation laws of $\aT^+$.
We note that the asymptotic correlation decay in non-integrable cases, say QC and NE, seems quite insensitive to 
increasing truncation order $n$ - indicating that the leading eigenvalues of $\aT^+_{[0,n]}$ remain frozen when 
increasing $n$. Note that the asymptotic exponents of correlation decay for a semi-infinite chain are not the same
as for an infinite one, i.e. the point spectra of $\aT^+_{[0,n]}$ and $\bT_n$ are in general different, however we have some
indications to believe that their phase diagrams should agree, namely ergodic regimes in the KI model on semi-infinite chain model are in one-to-one correspondence with ergodic regimes of the model on infinite chain.

However, the most remarkable feature of dynamics $\aT^+$ is the following. Splitting the truncated semi-lattice as 
$[0,n]=[0,m-1]\cup[m,n]$ and following the time evolution of an observable $A(t) = [\aT^+_{[0,n]}]^t A(0) = 
\sum_{\ve{s}} a^{(t)}_{(s_{0},\ldots s_n)} \Ket{s_0,\ldots,s_n}$
in terms of a `super-wave-function' $a^{(t)}_{(s_{0},\ldots,s_n)}$, 
one can again define the reduced super-density matrix as
\begin{equation}
R^{(m,n)}_{(s_{m},\ldots s_n),(s'_{m},\ldots,s'_n)}(t) = 
\!\!\!\sum_{s_{0},\ldots,s_{m-1}}\!\!\! a^{(t)}_{(s_{0},\dots,s_{m-1},s_{m},\ldots s_n)} a^{(t)*}_{(s_{0},\dots,s_{m-1},s'_{m},\ldots s'_n)}.
\label{eq:Rmatrixplus}
\end{equation}
The dynamical equation for $R^{(m+1,n+1)}(t+1)$ in terms of $R^{(m,n)}(t)$ 
is {\em exactly the same} as for doubly-infinite lattice, namely eqs. (\ref{eq:rdynamics1},\ref{eq:rdynamics2}).
Hence also the conjecture (\ref{eq:scaling}) on scaling invariance of $R^{(m,\infty)}(\infty)$ should be the same for the
two lattice topologies. However, numerical computations are much easier and thus the results are more suggestive for the 
semi-infinite case.
   
\begin{figure}
       \centering
       \includegraphics[width=16.0cm]{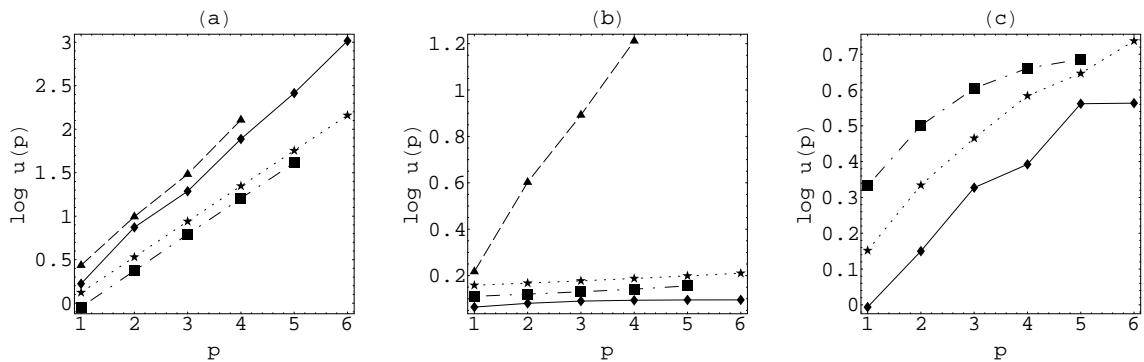}
       \caption{Scaling of the partial norms $u(p)=R^{(m_0+p,n)}_{0,0}(t)$ in log scale (units are arbitrary)
       versus the partial tracing index $p$, for cases QC (a), NE (b) and IN (c).
       Diamonds, stars, and squares represent data for the semi-infinite KI lattice with truncation
       size $n=15$,  (diamonds, stars) and $n=11$ (squares), all for $m_0=4$.
       Finite number of time steps $t=t^*=17$ (diamonds), just before the absorbing boundary affects any of 
       the data shown, is compared to steady state observable $t=\infty$ (stars).
       In the non-integrable cases (a,b), data are compared also with steady state $t=\infty$ simulation 
       of the same local initial observable $A(0) = \s{\x}{0}$ on the two-sided (doubly-infinite) KI 
       lattice with truncation size $n=7$ ($r=15$), and $m_0=2$ (triangles).
         }
       \label{fig:scalsinorm}
\end{figure}

Let us now discuss some numerical results. We have always started from local initial operator $A(0) = \s{\x}{0}$.
We took $n$ as large as allowed by existing computing resources, namely $n=15$ for the semi-infinite chain and
$n=7$ for the infinite chain, and compared the data for asymptotic matrices $R^{(m,n)}(\infty)$ 
(in numerics $t$ has been chosen such that the
results converged, typically $t\approx 100$) with finite time data 
$R^{(m,n)}(t^*)$ where time $t^*$ was set as large as
allowed so that the data were still exact and no truncation was needed, typically $t^* \approx n$.
In all cases, numerical results were quite insensitive to small changes in truncation order $n$.
First we have computed the scaling of the principal matrix element, or the {\em partial norms}
$R^{(m,n)}_{00}(t) = \sum_{\ldots s_{m-2},s_{m-1}\in\Z_4} |a^{(t)}_{(\ldots,s_{m-2},s_{m-1},0,0\ldots)}|^2$
which, assuming (\ref{eq:scaling}), should asymptotically scale as $\propto \zeta^m$ (see fig.\ref{fig:scalsinorm}).
If asymptotic dynamics $t\to\infty$ is determined by normalizable eigenvectors of $\aT^+$, which necessarily 
correspond to uni-modular eigenvalues, then we should have $\zeta = 1$.
In QC case a clear scaling was observed for both topologies ($\Z$ and $\Z^+$) with the
same exponent $\zeta$, however the exponent was slightly different for $R^{(m,n)}(\infty)$
and $R^{(m,n)}(t^*)$.
In the cases of non-ergodic dynamics (NE and IN) the results for two topologies were quite different.
For semi-infinite topology we find very clearly that $\zeta=1$ indicating that $A(t^*)$ and even $A(\infty)$
can be written as $l^2$ convergent sums of local operators.

 \begin{figure}
       \centering
       \includegraphics[width=16.0cm]{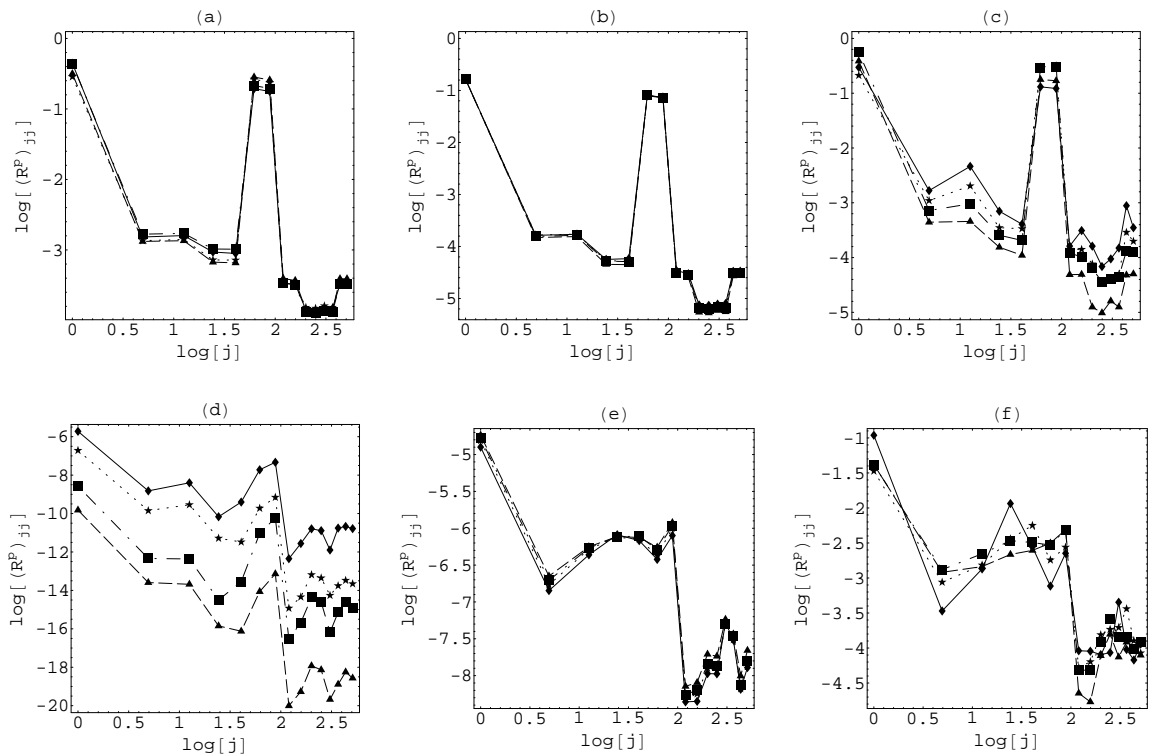}
       \caption{Scaling of the diagonal elements of scaled reduced super-density matrices 
       $R^{p}_{j,j}(t)$ 
       for initial tracing index $m_0=6$, and $p=1$ (diamonds), $p=2$ (stars), $p=3$ (squares), 
       $p=4$ (triangles), for 
       non-integrable cases of KI chain on semi-infinite lattice truncated at $n=15$, namely
       for case QC (a,b), and case NE (d,e).
       Finite time data at $t=t^*=17$ kicks are shown in (a,d), while steady state observables $t=\infty$ are 
       analyzed in
       (b,e), all starting from initial observable $\s{\x}{0}$.
       For comparison, asymptotic steady state $t=\infty$ data for two-sided, doubly infinite KI chain,  
       truncated at $r=13$, and with initial tracing index $m_0=2$, are shown in (c,f), namely for QC case 
       (c) and NE case (f).
       Note that data in plot (b) are practically exactly overlapping.}
        \label{fig:scalsi1d}
\end{figure}

\begin{figure}
       \centering
       \includegraphics[width=16.0cm]{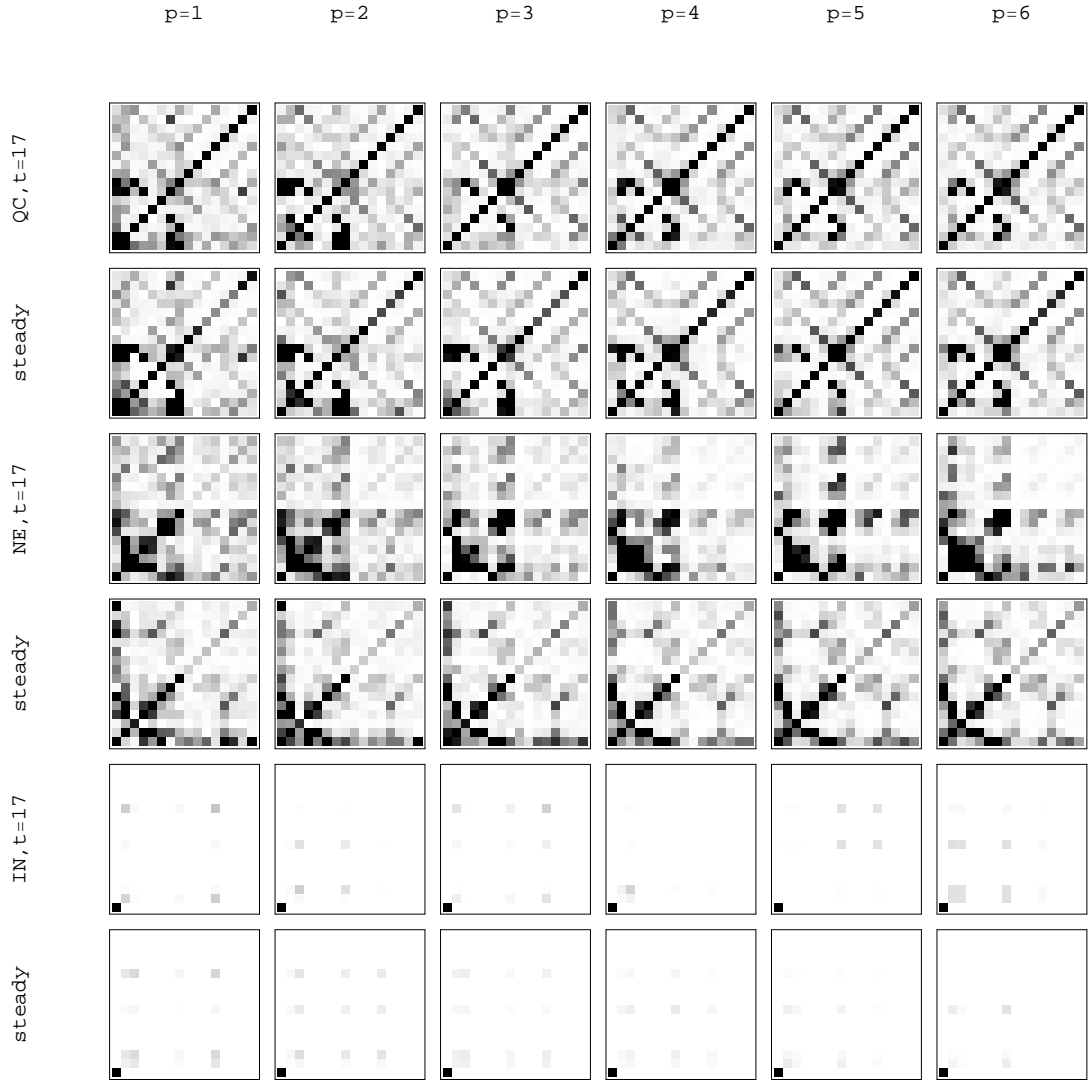}
       \caption{Scaling of scaled reduced super density matrices $R^{p}_{j,k}(t)$ - the grayness level is proportional
       to $|R^{p}_{j,k}(t)|$ -  for initial tracing index $m_0=4$, and  $p=1,\ldots,6$ (columns plots), 
       for six different cases, QC, NE, and IN of truncated semi-infinite
       KI lattice ($n=15$),
       at finite time $t=t^*=17$, and asymptotic steady state $t=\infty$ (row plots), always 
       starting form initial observable $\s{\x}{0}$.
       $16\times 16$ `most important' matrix elements are plotted with the matrix site $j=k=0$ at lower-left corner.}
         \label{fig:scalsi2d}
\end{figure}

As a more quantitative test of conjecture (\ref{eq:scaling}) we compare the upper-left (`most important') 
$16\times 16$ block of the super density matrix scaled to a unit principal element
$R^{p}_{j,k}(t) \equiv R^{(m_0+p,n)}_{j,k}(t)/R^{(m_0+p,n)}_{0,0}(t)$.
In fig.\ref{fig:scalsi1d} we plot the diagonal elements $R^{p}_{j,j}(t)$ for different $p$, while
in fig.\ref{fig:scalsi2d} we plot 2D charts of the entire scaled density matrices $R^{p}_{j,k}(t)$.
Indeed we find for QC case that the matrix $R^{p}_{j,k}(t)$ is practically insensitive to
increasing truncation ($p$) and to topology of the lattice (semi-infinite versus infinite),
both for finite time $t=t^*$ and `steady-state' $t=\infty$.
On the other hand, in non-ergodic cases, the scaling (\ref{eq:scaling}) is typically broken.
However, it seems to be observed in the steady state ($t=t^*$) of NE case, which is (in our setting) probably 
a non-physical but still quite robust effect due to truncation (a kind of absorbing boundary condition).

Summarizing this subsection, we conjectured that in the regime of quantum chaos
reduced super-density matrices of time-evolved observables typically obey the scaling law
(\ref{eq:scaling}) with the exponent $\zeta$ which only depends on global dynamics on quasi-local
algebra of observables and not on a particular choice of initial observable.
We expect that accurate numerical calculations of exponent $\zeta$ would be possible
within a certain quantum dynamic renormalization group scheme, however its precise
formulation at present remains an open problem.

\subsection{Dynamical entropies}

In section \ref{sect:MPS}, further elaborated in subsection \ref{sect:opent}, we have proposed
an entanglement in operator space of quasi-local algebra as a possible new measure of quantum
algorithmic complexity. Here we would like to compare this briefly to some 
established proposals of quantum dynamical entropy, such as for example CNT entropy \cite{CNT}
or AF entropy \cite{AF}, both being ideally suited for a quantum dynamical system formulated
in terms of time automorphism $\aT$, tracial invariant state $\omega$, and quasi-local 
$C^*$ algebra ${\cal A}_\Z$. We shall here briefly review AF entropy which is conceptually simpler.

\begin{figure}
       \vspace{1.5cm}
       \qquad\qquad\qquad\includegraphics[width=16.0cm,angle=-90]{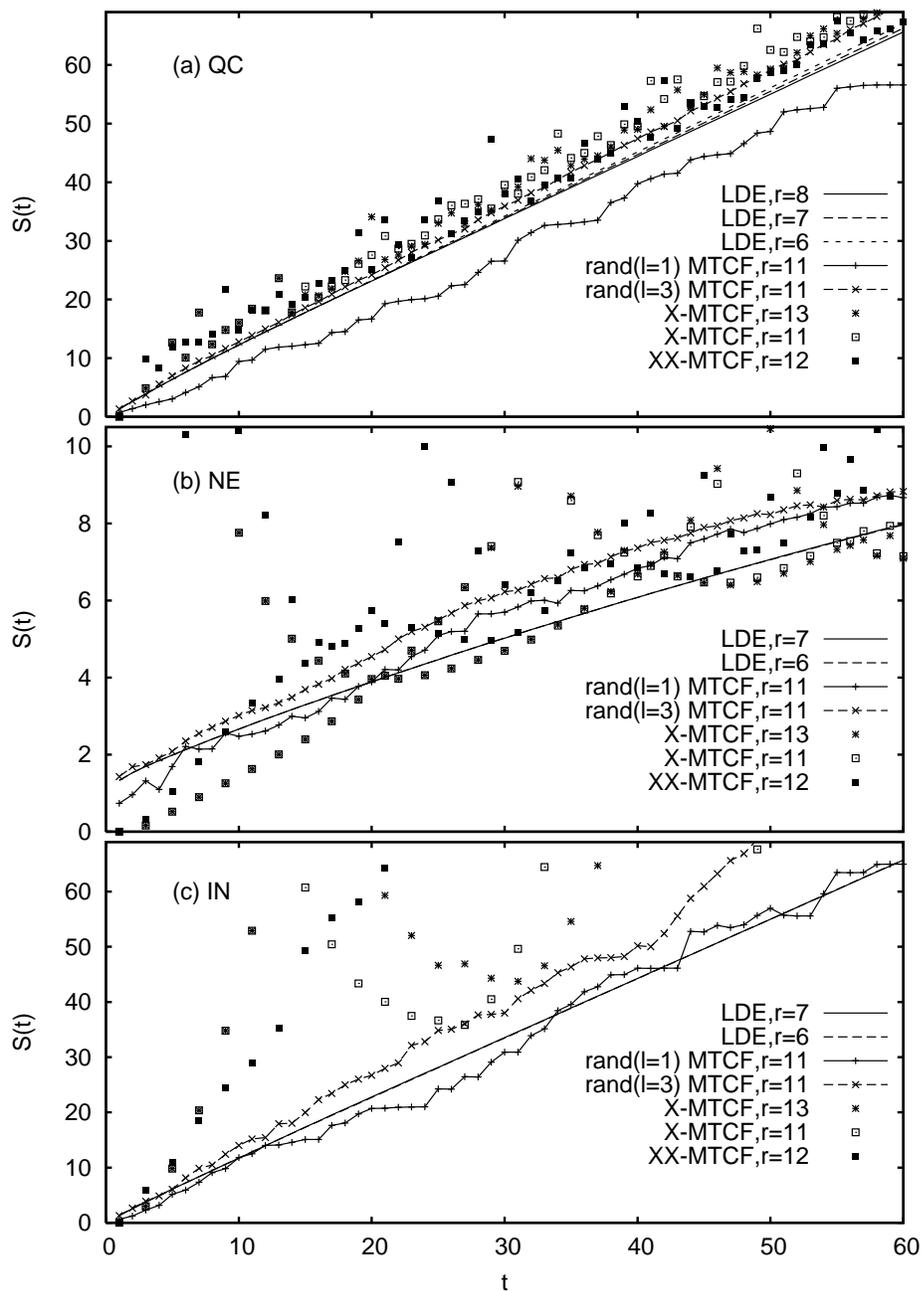}       
       \caption{Linear dynamical entropies (continuous curves) for cutoff orders at $r=6,7,8$, for three different
       cases of dynamics (QC (a), NE (b) and IN (c)), compared to $-\ln|C(t)|^2$ where $C(t)$ are MTCF with 
       obervables $X=\sigma^{\rm x}_1$
       and $XX = \sigma^{\rm x}_1 \sigma^{\rm x}_2$ at different cutoff orders $r$ (symbols), and to average
       $-\ln \ave{|C(t)|^2}$ over random MTCF
       sampled over $20$ sequences of random Pauli observables of lengths $l=1$ and $l=3$ (curve-symbols).}
       \label{fig:smb}
\end{figure}

Let us start by taking a set of say $k$ elements $A_\alpha,\alpha=1,\ldots k$ of ${\cal A}_\Z$ which form a partition of unity,
namely $\sum_\alpha A^*_\alpha A_\alpha = \one$.
Following the dynamics up to integer time $t$, $A_\alpha(t) = \aT^{t} A_\alpha$, 
we construct a set of $k^t$ elements depending on a multi-index 
$\ve{\alpha}=(\alpha_0,\ldots,\alpha_{t-1})$, namely 
\begin{equation}
A^{(t)}_{\ve{\alpha}} = A_{\alpha_0}(0)A_{\alpha_1}(1)\cdots A_{\alpha_{t-1}}(t-1) = 
A_{\alpha_0} \aT A_{\alpha_1} \aT \cdots  \aT A_{\alpha_{t-1}}.
\end{equation}
From the homomorphism property and unitality of dynamics it follows that $A^{(t)}_{\ve{\alpha}}$ is also a partition of unity, 
$\sum_{\ve{\alpha}} [A^{(t)}_{\ve{\alpha}}]^* A^{(t)}_{\ve{\alpha}} = \one $.
Hence using an invariant state $\omega$ one can form a {\em positive}, {\em Hermitian}, {\em trace-one},
$k^t \times k^t$ matrix
\begin{equation}
\rho^{(t)}_{\ve{\alpha},\ve{\beta}} = \omega\left( [A^{(t)}_{\ve{\beta}}]^* A^{(t)}_{\ve{\alpha}}\right).
\end{equation}
$\rho^{(t)}$ can clearly be interpreted as a density matrix pertaining to dynamically generated partition,
and its Von Neumann entropy generated per unit time defines
the AF entropy
\begin{equation}
S^{\rm AF} = \sup_{\{A_\alpha\}}\limsup_{t\to\infty} -\frac{1}{t}\tr\left[ \rho^{(t)} \log \rho^{(t)} \right],
\end{equation}
where, strictly speaking, supremum over all possible generating partitions has to be taken.
It is not surprising that practical evaluation of $S_{\rm AF}$ is impossible except for rather trivial cases, such
as  dynamics generated by shift automorphism $\aS_1$ \cite{AF}.
However, one can easily show that a related {\em linear AF entropy}
\begin{equation}
S^{\rm AF}_2 = \sup_{\{A_\alpha\}}\limsup_{t\to\infty} -\frac{1}{t}\log \tr [\rho^{(t)}]^2,
\label{eq:AFlin}
\end{equation}
is tractable much more easily, while the behaviour of $S^{\rm AF}$ and
$S^{\rm AF}_2$ is likely to be similar in practice.
The key observation is to write the purity
$P^{\rm AF}(t) = \tr[\rho^{(t)}]^2 = 
\sum_{\ve{\alpha},\ve{\beta}} \rho^{(t)}_{\ve{\beta},\ve{\alpha}} \rho^{(t)}_{\ve{\alpha},\ve{\beta}}$
in terms of dynamics over a product algebra
$\tilde{\cal A}_{\Z} = {\cal A}_\Z \times {\cal A}_\Z$, 
with time automorphism
$\cT(A\times B) = \aT(A)\times \aT(B)$
and an invariant state
$\tilde{\omega}(A\times B) = \omega(A)\omega(B)$. To this product structure we add a linear map 
$\cK :\tilde{\cal A}_{\Z} \to \tilde{\cal A}_{\Z}$ depending on a generating partition $\{A_\alpha\}$
\begin{equation}
\cK(\tilde{B}) = \sum_{\alpha,\beta=1}^k (A_\beta \times A_\alpha) \tilde{B} (A^*_\alpha\times A^*_\beta).
\end{equation}
If $\one$ is a unit element in ${\cal A}_\Z$ then $\one\times\one$ is a unit element in $\tilde{\cal A}_\Z$, and
purity can be written using a transfer-matrix-like approach as
\begin{equation}
P^{\rm AF}(t) = \tilde{\omega}\left\{ [\cT \cK]^t (\one \times \one)\right\} .
\end{equation} 
The idea can be worked out in detail for a {\em complete} generating partition of a local sub-algebra 
${\cal A}_{[-q+1,q]}$ of size $k=4^{2q}$. There it turns out that, due to locality of dynamics (\ref{eq:locality}), the resulting 
purity is {\em independent} of $q\ge 1$, i.e. the supremum (\ref{eq:AFlin}) is already achieved by a rather modest partition of $4^2=16$ elements, 
and the map $\cT \cK$ can be {\em factored} into a direct product of two maps acting separately on
{\em two independent copies} of a {\em semi-infinite lattice}. Leaving out some technical details of derivation, the final result reads as follows.
Let us consider a time dependent $4^t \times 4^t$ matrix $Q(t)$, representing a state
on $\tilde{\cal A}_{\Z^+} = {\cal A}_{\Z^+}\times {\cal A}_{\Z^+}$, with initial value $Q(0) = Q_{0,0}(0) = 1$,
and dynamics given by the following completely positive matrix map 
\begin{equation}
Q(t+1) = \tr_{[0,1]}\left\{T_{t+1}  \left[ E_{00}\otimes \one_4\otimes Q(t) \otimes E_{00} \right] 
T^\dagger_{t+1}\right\},
\label{eq:qdynamics}
\end{equation} 
where $(\tr_{[0,1]} R)_{j,k} = \sum_{s=0}^{15} R_{s + 16j,s+16k} $ traces out two least important qudits,
and unitary time evolution matrix $T_t$ is given in (\ref{eq:rdynamics2}). 
Then the purity, and linear dynamical entropy (LDE), are simply given as $P^{\rm AF}(t) = [Q_{0,0}(t)]^2$, 
$S_2(t) = -2 \log Q_{0,0}(t)$, respectively. The asymptotic increase of $S_2(t)$ per unit time yields the 
linear AF entropy (\ref{eq:AFlin}).
Again, for practical calculations it is convenient to consider truncation of matrices $Q(t)$ after each iteration 
(\ref{eq:qdynamics}), say at dimension $4^r$.
In fig.~\ref{fig:smb} we plot LDE for different cases of KI dynamics, and we observe that LDE is always clearly
growing linearly $\propto t$, so the AF entropy is always strictly positive, even in non-ergodic (NE) and integrable (IN) cases, and that the results are robust and stable against changing the truncation order $r$.
Perhaps this finding appears surprising, but one has to bear in mind that AF and CNT entropies can be positive
even for simpler dynamics, such as quasi-free flows on $C^*$ algebras.

For {\em ergodic} dynamical systems on $C^*$ algebras one can use Shannon-Mcmillan-Breiman (SMB)
theorem (for classical SMB theorem see e.g. \cite{biling}, and \cite{bjelakovic} for a possible quantum extension), which states that for {\em typical} sequences $\ve{\alpha}$, {\em multi-time correlation function} 
(MTCF) should decay exponentially 
\begin{equation}
C(t) = \omega(A^{(t)}_{\ve{\alpha}}) \sim \exp(-h t)
\label{eq:MCTF}
\end{equation}
where the exponent $h$, which should be essentially independent of $\alpha$, is equal to a dynamical entropy of the 
map $\aT$ with respect to an invariant state $\omega$.  For an interesting application of SMB theorem in the
context of quantum dynamical chaos see Ref.~\cite{Gaspard92}.
In our numerical experiments we considered truncated dynamics $\aT_{[-n,n]}$, writing truncation order as
$r=2n+1$, and computed two kinds of MTCF: (i)
For a uniform sequence $\ve{\alpha}=(1,1,\ldots)$, where $A_1=\s{\x}{0}$ (or $A_1=\s{\x}{0}\s{\x}{1}$ in which
case the truncated lattice was placed as $[-n,n+1]$, hence $r=2n+2$), we estimated MTCF (\ref{eq:MCTF}) as
$ \Braket{A_1}{\aT_{[-n,n]} A_1 \aT_{[-n,n]}  \cdots \aT_{[-n,n]} A_1}$.
(ii) For a random sequence $\ve{\alpha}$, we took a complete
generating partition $\{ \Ket{s_{-q},\ldots s_{q}} \}$ on ${\cal A}_{[-q,q]}$ with $k=4^l$ elements, for $l\equiv 2q+1=1$ or $l=3$,
and computed an average square modulus of MTCF
$\ave{|C(t)|^2} = \ave{|\Braket{A_{\alpha_0}}{
\aT_{[-n,n]} A_{\alpha_1} \aT_{[-n,n]}  \cdots \aT_{[-n,n]} A_{\alpha_{t-1}}}|^2}$
over 20 randomly sampled sequences $\ve{\alpha}$.
In both cases (i,ii) we have found a rather good agreement between $\log |C(t)|^2$ and LDE $S_2(t)$
for the case of quantum chaotic dynamics (fig.~\ref{fig:smb}). 
For average random  MTCF we found rather good agreement even for non-ergodic cases (NE and IN), however
MTCF for the uniform sequence exhibited big oscillating fluctuations there, indicating simply that SMB theorem does hold
for non-ergodic dynamics of KI model.

Summarizing, it seems that AF entropy, even though it very cleanly generalizes the concept of
Kolmogorov-Sinai classical dynamical entropy to quantum dynamical systems, 
cannot be used as an indicator of quantum chaos or an indicator of computational complexity of quantum 
dynamical systems. Nevertheless, it has to be stressed that quantum dynamical entropies can
be related to the notion of quantum algorithmic complexity, mentioned in section \ref{sect:MPS},
by a quantum version of Brudno theorem established in Ref. \cite{benattietal}.
 
\section{Conclusions and open questions}

\label{sect:conc}

In the present article we have reviewed several possible approaches to describe dynamic instabilities, relaxation phenomena, and computation complexity in the simple model of one-dimensional non-integrable locally
interacting quantum many body dynamics. We have argued that the essential non-equilibrium statistical properties of quantum dynamical systems - in the absence of idealized external baths - may be crucially related to the integrability of the system, or in the complementary case, to the existence of the regime of quantum chaos.

The general flavor which remains after such studies is that the non-integrable quantum many body problem at high temperature will
preclude any exact and complete solution by its very nature. Still it is hoped that a more complete ergodic theory of
such systems could be developed, allowing for example for exact calculation of relaxation rates, scaling exponents of resonance
eigenvectors, etc.

It has been shown in many cases that integrable quantum systems have anomalous non-equilibrium statistical mechanics at high temperature, 
for example they exhibit ballistic transport. This can easily be understood as being a consequence of existence of (an infinite sequence of) exact local conservation laws which prevents quantum ergodicity, similarly as existence of canonical action variables in classical integrable systems prevents classical ergodicity.
An interesting open question is the following: How strong integrability breaking perturbation is needed, in generic cases, to break all the exact conservation laws and yield normal (diffusive) transport? In other words: A quantum KAM theory is needed! Numerical experiments shown in this 
paper suggest an interesting possibility, namely that in some cases a finite, critical perturbation strength is required. 

\section*{Acknowledgements}

I would like to thank G. Casati, T. Gorin, C. Mejia-Monasterio, T. H. Seligman, and M. \v Znidari\v c, for collaborations 
on related projects and acknowledge stimulating discussions with F. Benatti, J. Eisert, S. Fishman and P. Zoller.
This work has been supported by the grants P1-0044, and J1-7347 of Slovenian Research Agency.

\end{document}